\documentclass[usenatbib]{mn2e}
\usepackage{color}
\usepackage{natbib}
\usepackage{paralist}
\usepackage{graphicx}
\usepackage{epstopdf}
\usepackage{epsfig}
\usepackage{amsmath}
\usepackage{tablefootnote}

\usepackage{amssymb}
\usepackage{lipsum}
\usepackage{float}
\usepackage{array}
\usepackage{multirow}
\usepackage{setspace}

\title[Mapping the aliphatic hydrocarbon content of ISD]{Mapping the aliphatic hydrocarbon content of interstellar dust in the Galactic plane}
\author[B. G\"{u}nay, M. G. Burton, M. Af\c{s}ar, T. W. Schmidt]
{B. G\"{u}nay $^{1,3}$, M. G. Burton$^{2,3}$, M. Af\c{s}ar$^{1}$, T. W. Schmidt$^{4}$ \\
\\
\thanks{E-mail: burcu.gunay@ege.edu.tr (BG)}
$^{1}$Department of Astronomy and Space Sciences,
                 Ege University, 35100 Bornova, \.{I}zmir, Turkey\\
$^{2}$Armagh Observatory and Planetarium,
                 College Hill, Armagh, BT61 9DG, Northern Ireland, UK\\
$^{3}$School of Physics,
                 UNSW Sydney, NSW 2052, Australia\\
$^{4}$ARC Centre of Excellence in Exciton Science, School of Chemistry,
                 UNSW Sydney, NSW 2052, Australia\\
}

\begin{document}

\maketitle
\label{firstpage}
\begin{abstract}
We implement a new observational method for mapping the aliphatic hydrocarbon content in the solid phase in our Galaxy, based on spectrophotometric imaging of the 3.4\,$\mu$m absorption feature from interstellar dust. We previously demonstrated this method in a field including the Galactic Centre cluster. 
We applied the method to a new field in the Galactic centre where the 3.4\,$\mu$m absorption feature has not been previously measured and we extended the measurements to a field in the Galactic plane to sample the diffuse local interstellar medium, where the 3.4\,$\mu$m absorption feature has been previously measured. We have analysed 3.4\,$\mu$m optical depth and aliphatic hydrocarbon column density maps for these fields.
Optical depths are found to be reasonably uniform in each field, without large source-to-source variations. There is, however, a weak trend towards increasing optical depth in a direction towards $b=0^{\circ}$ in the Galactic centre. 
The mean value of column densities and abundances for aliphatic hydrocarbon were found to be about several $\rm \times 10^{18} \, cm^{-2}$ and several tens $\times 10^{-6}$, respectively for the new sightlines in the Galactic plane. We conclude that at least 10--20\% of the carbon in the Galactic plane lies in aliphatic form.

\end{abstract}


\section{Introduction}\label{Section1}

Carbon is a key element in the evolution of organic material in the Universe. There is a rich carbon chemistry in our Galaxy and other galaxies because of its abundance and its ability to form complex molecules. Carbon chemistry starts in the circumstellar medium of evolved massive stars \citep{Henning1998}. It branches through different phases of the interstellar medium (ISM)\@. The material cycle between the stars and the gas in the ISM leads to the delivery of organic molecules in molecular clouds and planetary systems. A special sub-class of organic molecules called prebiotic molecules are thought to play a major role in the formation of life on Earth \citep{ChybaSagan1992, Ehrenfreund2010, Ehrenfreund2000}. Therefore, it is important to understand the life cycle of organic molecules and of the element carbon in space.

Carbon is the $4^{th}$ most abundant element in the Universe. The elemental carbon abundance is the total carbon abundance in the gas and solid phases of the ISM. The total carbon abundance observed in the ISM should be in agreement with the cosmic carbon abundance estimations \citep{Zuo2021a, Zuo2021b, HensleyDrain2021}.

The cosmic carbon abundance derived from the Solar atmosphere \citep{Grevesse1998, Turcotte2002, Chiappini2003, Asplund2005, Asplund2009, Asplund2021}, the Solar System abundances \citep{Lodders2003, Asplund2021}, Sun-like stars \citep{Bedell2018}, the atmosphere of young stars \citep{SnowWitt1995, Sofia2001, Bensby2006, Przybilla2008} imply that, there is up to $\sim$$358$\,ppm\footnote{ppm: parts per million} of carbon available in the ISM.

Carbon is one of the major dust-forming elements. The cosmic carbon abundance sets a limit to the maximum amount of carbonaceous material available for making the interstellar dust (ISD) that accounts for the observed extinction, so that carbon abundance hidden in the ISD should be in accordance with the available carbon and other dust-forming elements in the ISM \citep{Zuo2021a, Zuo2021b, HensleyDrain2021}. The depletion of carbon in the gas phase refers to its missing part with respect to cosmic abundance. However, carbon depletion from the gas phase is not sufficient to account for the observed extinction. This has been called the \textit{carbon crisis} \citep{Kim1996, Cardelli1996, Henning2004, Wang2015, Zuo2021b}. 

The wavelength dependence of the extinction (A$_{\lambda}$) can be defined by an extinction curve. The interstellar extinction curves give clues as to the size and chemical composition of the dust particles \citep{Cardelli1989, Fitzpatrick1999}.  The regimes covering the far-ultraviolet (UV) and mid-infrared (MIR) exhibit features of the main components of dust, carbonaceous and siliceous materials \citep{DraineISD2003, Gordon2019, Gordon2021, HensleyDrain2021}. Studies of the extinction curves show that they are spatially variable through the Galaxy. The size, structure and chemical composition of the ISD play a role in this variability \citep{Fitzpatrick1999, Fitzpatrick2019, Zuo2021b}. Since the composition and structure of dust is variable, \citep{Henning2004, Jones2012a, Jones2012b, Jones2012c, Jones2013, Jones2019, DraineHensley2021}, elemental abundance estimations taking into account particle size to reproduce extinction curves are prone to large discrepancies \citep{Mathis1977, Draine1984, Kim1996, Mathis1996, Li1997, Mishra2017, Zuo2021b}.

Therefore, an observational method to trace carbonaceous ISD is required, as this invisible solid component is a reservoir for organic material and the element carbon \citep{Sandford1991, VanDishoeck2014}.

Carbon molecules in the gas phase can be studied using the electromagnetic spectrum from ultraviolet to radio frequencies. Although complex molecules can be discerned in diffuse molecular gas through their vibrational and rotational spectra, it is more challenging to study large molecules, as their spectra are complex \citep{McGuire2018}. In particular, when they are in the solid phase, only vibrational spectra can be used for searching for some chemical groups and for probing the carbonaceous material in the ISD\@. 

There are some useful emission and absorption features in the infrared spectral region for this purpose. The 3.4\,$\mu$m (2940 cm$^{-1}$) absorption feature is of particular interest since it is prominent and suitable for observational measurements. This feature arises due to the aliphatic C--H stretch of methylene (--CH$_{2}$--) and methyl (--CH$_{3}$) groups in carbonaceous material in the ISD\@. The strength of the 3.4\,$\mu$m absorption feature is proportional to the number of aliphatic C--H bonds. To estimate the amount of aliphatic hydrocarbons in solid phase in the ISM, measurements of the 3.4\,$\mu$m absorption from ISD can be combined with the absorption coefficient measurements of interstellar dust analogues (ISDAs) produced in a laboratory. 
	
We undertook laboratory measurements in order to provide a revised value for the absorption coefficient of aliphatic hydrocarbons (\citealt{Gunay2018}, hereafter Paper 1). We produced ISDAs under simulated circumstellar / interstellar-like conditions. The integrated absorption coefficient ($A$, cm molecule$^{-1}$) and line width ($\Delta \bar{\nu}$, cm$^{-1}$) (for the low-resolution spectra it becomes the filter bandwidth) that were measured from ISDAs and the optical depth ($\tau$) of the 3.4\,$\mu$m absorption can be used to obtain column density (N, cm$^{-2}$) of aliphatic hydrocarbon groups, as follows: 

\begin{equation}\label{eq:1}
N =\frac{\tau \Delta \bar{\nu}} {A}
\end{equation}
				
In order to investigate the amount and distribution of aliphatic hydrocarbon abundances incorporated in the ISD in our Galaxy we need to map the 3.4\,$\mu$m optical depth, $\tau_{3.4\,\mu m}$, over wide fields. This requires us to obtain optical depth at 3.4\,$\mu$m for as many sightlines as possible. Measurement of the optical depth can be done readily for individual sightlines by single point or long-slit spectroscopy. As these spectroscopic processes are comparatively slow and require long observing times, Integral Field Spectroscopy (IFS) \citep{Allington2006} is used to speed up observations by simultaneously obtaining spectra in a two-dimensional field. It has become an important method in cases where there is a need to examine the spectra of extended objects (such as the ISM) as a function of position. However, narrow-band imaging can also be used to obtain spatially resolved spectral information for larger fields of view.

We previously implemented a new method  (\citealt{Gunay2020}, hereafter Paper 2) and trialed it through the dusty sightlines of the diffuse interstellar medium (DISM) towards a field that contains the centre of the Galaxy (hereafter Field A), where the optical depth of the 3.4\,$\mu$m absorption had already been reported in the literature (references in Paper 2). We did this in order to test the veracity of this new method, using narrow-band filters spread across the absorption feature.  Several well-studied bright sources in the field were used whose data were available in NASA/IPAC Infrared Science Archive (IRSA\footnote{https://irsa.ipac.caltech.edu/frontpage/}). Their brightness values are listed in Spitzer\footnote{https://irsa.ipac.caltech.edu/Missions/spitzer.html} and 2MASS\footnote{https://irsa.ipac.caltech.edu/Missions/2mass.html} catalogs. Their spectra were previously reported by \cite{Chiar2002} (hereafter [C02]) and \cite{Moultaka2004} (hereafter [M04]), and provided calibration. We used these to determine zero points for our measurements, and thence optical depths at 3.4\,$\mu$m (as described in detail in Paper 2).

We applied Equation~\ref{eq:1} using the 3.4\,$\mu$m narrow-band filter width (62\,cm$^{-1}$). We then calculated aliphatic hydrocarbon column densities using the absorption coefficient of (\textit{A} = $4.7\times10^{-18}$\,cm\,group$^{-1}$) from Paper 1. In demonstrating that the technique worked, we were able to produce an aliphatic hydrocarbon column density map for the Galactic Centre cluster in Paper 2.  We found a mean value of $\tau_{3.4\,\mu m}$ $\sim$ 0.2, corresponding to a typical aliphatic hydrocarbon column density of $\sim 3 \times10^{18}$\,cm$^{-2}$. Further, there were indications that the column density is increasing in a direction towards the Galactic mid-plane.

Normalised aliphatic carbon abundances in ppm were also calculated based on a gas-to-extinction ratio $N(H) = 2.04 \times 10^{21}$\,cm$^{-2}$  mag$^{-1}$ \citep{Zhu2017}, by assuming A$_{V}$$\sim$30 mag. A mean value of 43\,ppm aliphatic hydrocarbon abundance was found. Comparing this to the ISM total carbon abundance of 358\,ppm \citep{Sofia2001}, we found that approximately 12$\%$ of the carbon is in aliphatic form. This shows that ISD is an important reservoir for aliphatic hydrocarbons in the Galactic field studied. 

In this work, we have obtained maps of the 3.4\,$\mu$m optical depth across the two new fields and investigated the amount of aliphatic hydrocarbons. We compared the new maps with the maps previously obtained for Galactic Centre cluster field (Paper 2). We discuss the amount and distribution of aliphatic hydrocarbons in these three fields in the Galactic plane. We also tried to investigate whether there are similarities between the distribution of the aliphatic hydrocarbons in ISD and the total dust, for the Galactic Centre fields. However we could not find any relation.

This paper is organised as follows. The observations and data reduction are described in Section~\ref{Section2}, data analysis in Section~\ref{Section3}, mapping applications in Section~\ref{Section4}, results in Section~\ref{Section5}, and the discussion and conclusions are presented in Section~\ref{Section6}.

\section{Observations and Data Reduction}\label{Section2}

In this third paper we have now applied this new technique for mapping the 3.4\,$\mu$m optical depth to two other fields; one in the Galactic Centre region (hereafter Field B), but well separated from the well-studied central cluster, the other one in the Galactic plane (hereafter Field C), well away from the Galactic Centre. 

To the best of our knowledge, Field B has not yet been examined for the aliphatic hydrocarbon absorption feature. It has been chosen to contain several sources with brightness m$_{L}$ \textless 7$^{m}$ and many with m$_{L}$ \textless 10$^{m}$, which allow us to apply the method. The brightest 180 sources with m$_{L}$ \textless 10$^{m}$ were used for the spectrophotometric measurements.

Field C was chosen to be well away from the Galactic Centre to determine whether the method could be applied to the spiral arm regions of the Milky Way. Field C samples the DISM and is less affected by strong extinction, unlike the Galactic Centre region fields, which sample other more local extinction of the ISM of Galactic nucleus \citep{Moultaka2019, Geballe2021}. It has been chosen towards the IRAS 18511+0146 cluster, where the 3.4$\,\mu$m absorption was previously reported by \cite{Ishii2002} (hereafter [I02]) and \cite{Godard2012} (hereafter [G12]). We have applied the method by using the fluxes of two sources from the IRAS 18511+0146 cluster listed in [G12] and compared our measurements with the optical depths reported in [I02] and [G12]. The spectrophotometric observations in this study cover a larger field of view (137 arcsec) than the previous studies on IRAS 18511+0146 \citep{Vig2007, Vig2017} and include optical depth measurements of more sources than [I02] and [G12]. There are several L--band (3.6$\,\mu$m) sources with brightness of m$_{L}$ \textless 11$^{m}$ and photometric data available \citep{Vig2007, Vig2017}. We detected 18 sources and used 15 of them with m$_{L}$ \textless 11$^{m}$ for the spectrophotometric measurements.

Observations were carried out on the 3.8\,m United Kingdom Infrared Telescope (UKIRT) using the UIST camera\footnote{https://www.ukirt.hawaii.edu/instruments/uist/uist.html} ($1 - 5\,\mu$m imager--spectrograph, 1024 $\times$ 1024 InSb array, 0.12 arcsec /pixel with 123 arcsec field of view).  Data were obtained through service observations for two fields in the Galactic Centre region; Field A and Field B (Project ID: U/15B/D01, 15--25 September 2015) and additionally for one field in the Galactic plane; Field C (Project ID: U/16B/D01, 20--21 September 2016). The results for Field A (which includes the Galactic Centre itself) were reported in Paper 2, and the methodology used here follows that described in that paper. We summarise the parameters for the three fields in Table~\ref{tab:1}. 

Spectrophotometric measurements had been obtained using narrow-band filters at 3.05$\,\mu$m, 3.29$\,\mu$m, 3.4$\,\mu$m, 3.5$\,\mu$m, 3.6$\,\mu$m and 3.99$\,\mu$m. However, we used only measurements with three filter 3.29$\,\mu$m, 3.4$\,\mu$m, 3.6$\,\mu$m in order to measure the optical depth of the 3.4$\,\mu$m absorption feature, as we followed the same methodology in our previous work (Paper 2). We calculated the sensitivity thresholds (mag) for signal-to-noise ratio (S/N) = 5 using the UIST online calculator\footnote{http://www.ukirt.hawaii.edu/cgi-bin/ITC/itc.pl} for the total integration time we applied for each filter (see table~2 in Paper~2).

The fields were imaged with a 3$\times$3 jitter pattern and 1 minute of integration per jitter position.  The 9 point jitter pattern was then repeated, to achieve a total on-source integration time of 18 minutes.  While Fields A and B were observed with a 20 arcsec grid pattern, Field C used a 7 arcsec grid, being less crowded. The pixel scale was 0.12 arcsec. The resultant fields of view (FoV) were 163 arcsec for Fields A and B and 137 arcsec for Field C (see Table~\ref{tab:1}). The seeing ranged between 0.7 to 0.8 arcsec at 3.6$\,\mu$m during the observations.

\begin{table*}
 \begin{center}
  \caption{Parameters for the observed fields.}
  \label{tab:1}
   \centering
  \begin{tabular}{| c | c | c | c | }
  \hline
Targets & Field A  & Field B  & Field C  \\
  \hline

Longitude & $l: 359.945^{\circ}$ & $l: 359.765^{\circ}$ & $l: 034.821^{\circ}$ \\
Latitude & $b: -0.045^{\circ}$ & $b: -0.045^{\circ}$ & $b: 0.352^{\circ}$ \\
 \hline
Field of View&163 arcsec & 163 arcsec & 137 arcsec \\
 \hline
 \multirow{1}{*}{Jitter}&9pt jitter & 9pt jitter & 9pt jitter \\
 \multirow{1}{*}{Pattern}&offsets of  $20''$ & offsets of $20''$ & offsets of  $7''$\\
  \hline
Observing Dates & September 2015 & September 2015 & September 2016\\

 \hline
\end{tabular}
\end{center}
\end{table*}

\begin{figure*}
\begin{center}
    \begin{tabular}{ccc}
         {\includegraphics[trim=130mm 15mm 140mm 15mm, clip, angle=0,scale=0.55]{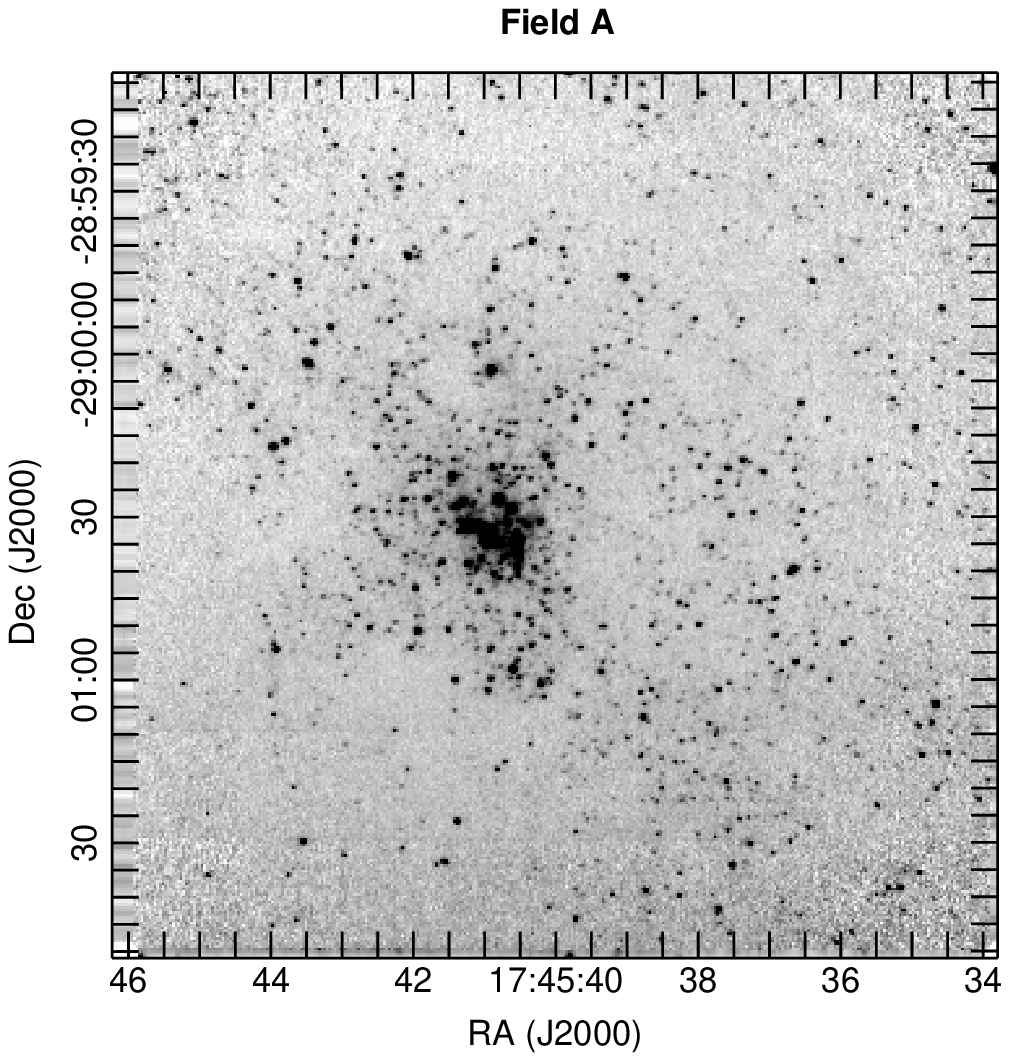}} &  {\includegraphics[trim=130mm 15mm 140mm 15mm, clip, angle=0,scale=0.55]{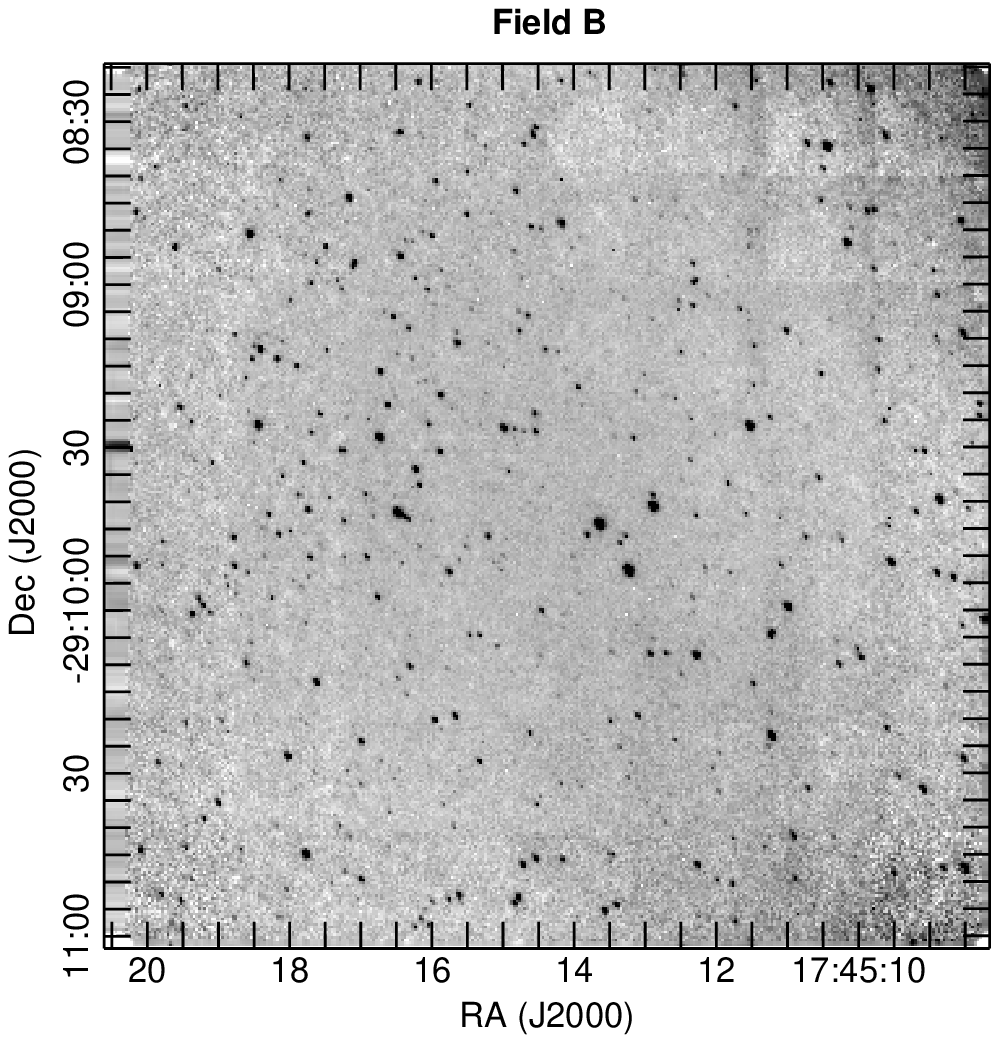}} &  {\includegraphics[trim=140mm 15mm 140mm 15mm, clip, angle=0,scale=0.55]{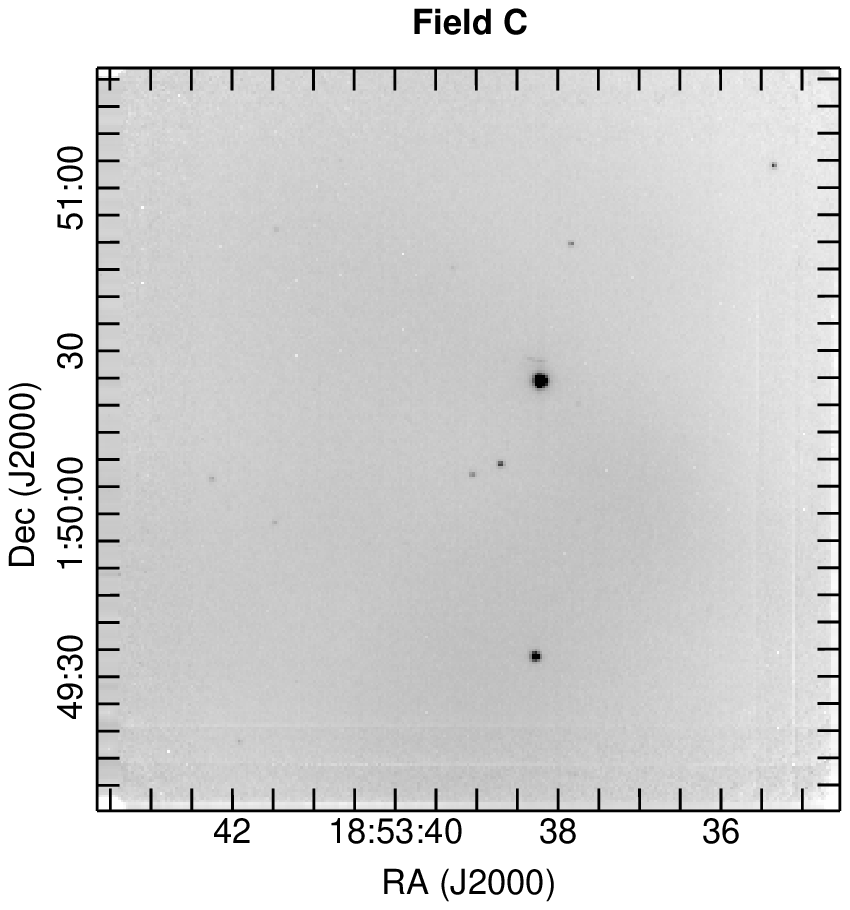}}\\  
    \end{tabular}
          \caption{Images of Fields A, B and C obtained through the 3.4$\,\mu$m filter. Coordinates are celestial (J2000).}      
          \label{fig1}
     \end{center}
\end{figure*}

Data reduction was carried out using the Starlink\footnote{http://starlink.eao.hawaii.edu/starlink} / ORAC-DR\footnote{http://www.starlink.ac.uk/docs/sun230.htx/sun230.html} data reduction pipeline using the recipe described in Paper 2. For each jittered frame a reduced image is created and added into a master mosaic to improve signal to noise. The final mosaics were not trimmed to the dimensions of a single frame, thus the noise is greater in the peripheral areas which have received less total exposure time. The resultant images for each of the three fields in the 3.4$\,\mu$m filter are shown in Figure~\ref{fig1}.

\section{Data Analysis}\label{Section3}

\subsection{Photometric measurements}

As per Paper 2, optimal photometry was carried out by using the Starlink / GAIA\footnote{http://star-www.dur.ac.uk/~pdraper/gaia/gaia.html} package, involving profile fitting of bright and isolated stars \citep{Naylor1998}.  

Using the sensitivity thresholds for each filter (see table~2 in Paper~2), we determined that the minimum brightness level of 11.9$^{m}$ is required for the measurements with 3.29$\,\mu$m filter, 12.2$^{m}$ is required for the 3.4$\,\mu$m filter and 12.1$^{m}$ is required for the 3.6$\,\mu$m filter to satisfy the S/N = 5 criterion. To ensure that the optical depth measurements are obtained with as highest S/N ratio as possible, we eliminated the stars fainter than 11.0$^{m}$. Previously, we analysed the brightest 200 sources with magnitudes of $m_{3.6\,\mu m}$ $\le$ 9.5$^{m}$ in Field A (Paper 2). For Field B the number of bright sources is a little less than in Field A and the brightest 180 sources with magnitudes $m_{3.6\,\mu m}$ $\le$ 10$^{m}$ are selected. For Field C, only 15 sources with magnitudes of $m_{3.6\,\mu m}$ $\le$ 11$^{m}$ are found to be useful.  

\subsection{Zero point calibration}

In the previous study, to calibrate Field A data, we extracted spectral fluxes for the bright L--band sources in the GC cluster, as measured by [C02] and [M04] to determine zero points (ZPs) for each filter. After the comparison, GCIRS 7 as measured by [C02] was chosen (hereafter referred to as GCIRS7--C02) to calibrate Field A data (described in detail in Paper 2). The fluxes for GCIRS7--C02 and the ZPs so obtained have been presented in Paper 2 (tables \ref{tab:3} and \ref{tab:4}). In this study, we applied the same ZP values to calibrate the narrow-band measurements of the new fields. This also allowed us to compare and interpret the optical depth measurements of Field B and Field C together with Field A.

Unlike Field A, where we previously compared our results with the values obtained by [M04], there are no sources within the Field B available to provide a self-consistency check on this calibration, since the optical depths at 3.4$\,\mu$m have not been measured before. In the case of Field C there are sources previously studied in the literature (\citealt{Vig2007, Vig2017}, [I02] and [G12]), so we used Field C for checking the cross-calibration procedure we applied.

We present the calibrated fluxes for the sources in Field C at 3.3$\,\mu$m, 3.4$\,\mu$m and 3.6$\,\mu$m as well as the derived 3.4\,$\mu$m optical depths, $\tau_{3.4\,\mu m}$, in Table \ref{tab:2}, together with their galactic and celestial coordinates. Their 2MASS K-band (2.17\,$\mu$m) band brightnesses and Spitzer IRAC Ch1(3.6\,$\mu$m) brightnesses are also listed and corresponding SSTGLMA\footnote{https://irsa.ipac.caltech.edu/data/SPITZER/GLIMPSE/} (Spitzer Space Telescope The GLIMPSE Archive) designations are indicated. For comparison, we also note the corresponding IDs of the three sources for which the 3.4\,$\mu$m optical depths were previously measured by [G12] (i.e.\  S7, S10, S11) in the footnote to Table \ref{tab:2}.

\begin{table*}
 \begin{center}
  \caption{Source IDs, galactic and celestial coordinates (J2000) in degrees, fluxes ($\times$\,10$^{-17}$ W\,cm$^{-2}$ $\mu$m$^{-1}$) and 3.4\,$\mu$m optical depths ($\tau_{3.4\,\mu m}$) determined for the sources in Field C. In addition, corresponding SSTGLMA designations, and brightness values (mag) from 2MASS at 2.17$\,\mu$m and Spitzer IRAC at 3.6$\,\mu$m, are also given}.
\centering
  \label{tab:2}
  \begin{tabular}{| c | c | c | c | c | c  c  c | c | c | c  c |}
    \hline

Source & \multirow{2}{*}{ \textit {l} }  & \multirow{2}{*}{ \textit {b} } & \multirow{2}{*}{ RA } &	\multirow{2}{*}{ Dec} &	\multicolumn{3}{c|}{Fluxes }  &	 \multirow{2}{*}{$\tau_{3.4\,\mu m}$}	&	\multirow{2}{*}{ SSTGLMA  } &  \multicolumn{2}{c|}{Brightnesses}  \\

No	 &   & & &	& 3.3$\,\mu$m	&	3.4$\,\mu$m &	 3.6$\,\mu$m	&		&	&2.17$\,\mu$m  &  3.6$\,\mu$m \\
    \hline
1	&	34.820	&	0.351	&	283.4091	&	1.8401	&	41.03	&	32.53	&	41.92	&	0.24	&	-	&	-	&	-	\\
2	&	34.808	&	0.344	&	283.4094	&	1.8261	&	8.36	&	6.37	&	6.20	&	0.18	&	G034.8087+00.3453	&	6.10	&	6.79	\\
3	&	34.817	&	0.347	&	283.4112	&	1.8359	&	0.82	&	0.82	&	1.19	&	0.14	&	G034.8183+00.3482	&	10.97	&	7.68	\\
4	&	34.824	&	0.366	&	283.3973	&	1.8511	&	1.04	&	0.67	&	0.86	&	0.39	&	-	&	-	&	-	\\
5	&	34.817	&	0.345	&	283.4126	&	1.8353	&	0.30	&	0.37	&	0.63	&	0.10	&	G034.8185+00.3467	&	12.66	&	7.75	\\
6	&	34.825	&	0.355	&	283.4076	&	1.8471	&	0.52	&	0.33	&	0.35	&	0.33	&	G034.8266+00.3565	&	9.25	&	9.01	\\
7	&	34.823	&	0.333	&	283.4259	&	1.8351	&	0.32	&	0.22	&	0.20	&	0.26	&	G034.8243+00.3348	&	9.76	&	9.58	\\
8	&	34.811	&	0.328	&	283.4245	&	1.8217	&	0.16	&	0.14	&	0.18	&	0.20	&	G034.8117+00.3299	&	12.27	&	9.93	\\
9	&	34.820	&	0.335	&	283.4227	&	1.8329	&	0.14	&	0.11	&	0.13	&	0.20	&	G034.8208+00.3366	&	11.95	&	10.14	\\
10	&	34.833	&	0.342	&	283.4226	&	1.8478	&	0.19	&	0.12	&	0.13	&	0.34	&	G034.8341+00.3435	&	10.44	&	10.22	\\
11	&	34.827	&	0.349	&	283.4136	&	1.8459	&	0.11	&	0.08	&	0.08	&	0.29	&	G034.8282+00.3506	&	11.02	&	10.61	\\
12	&	34.813	&	0.349	&	283.4073	&	1.8330	&	0.05	&	0.04	&	0.06	&	0.33	&	G034.8139+00.3504	&	13.11	&	10.95	\\
13	&	34.817	&	0.356	&	283.4027	&	1.8403	&	0.06	&	0.04	&	0.06	&	0.41	&	G034.8184+00.3578	&	13.17	&	11.07	\\
14	&	34.818	&	0.352	&	283.4072	&	1.8389	&	0.08	&	0.06	&	0.05	&	0.19	&	-	&	-	&	-	\\
15	&	34.810	&	0.335	&	283.4185	&	1.8240	&	0.05	&	0.03	&	0.04	&	0.44	&	G034.8110+00.3362	&	12.78	&	11.50	\\

\hline
\end{tabular}
  \begin{flushleft}
\begin{footnotesize}
Note that the IDs for the sources measured by [G12] are as follows:  1:S7, 3:S10, 5:S11.
\end{footnotesize}
\end{flushleft}

    \end{center}
  \end{table*}

\begin{figure*}
  \begin{center}
    \begin{tabular}{c}
       {\includegraphics[angle=0,scale=0.42]{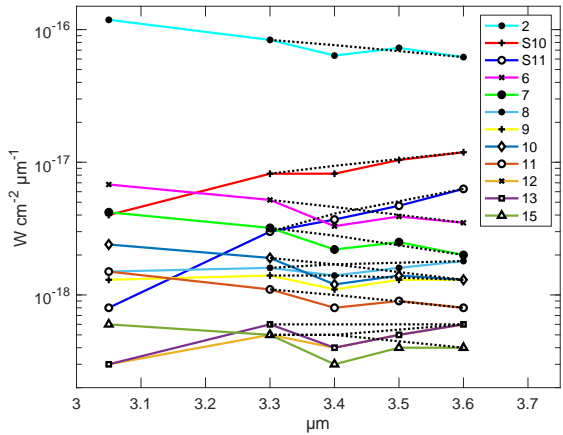}} 
                \end{tabular}

    \caption{The fluxes for Field C sources obtained by spectrophotometric calibration using GCIRS7 as described in the text. The 3.4$\,\mu$m optical depths were then calculated by interpolating across the aliphatic absorption feature from 3.3$\,\mu$m  to 3.6$\,\mu$m, as shown by the dotted lines. }
     \label{fig2}
    
      \end{center}
\end{figure*}

The calibrated spectra for the Field C sources are shown in Figure \ref{fig2} (on the left panel). They are compared with the flux levels (on the right panel) calculated using interpolation between 2MASS K and Spitzer IRAC Ch1 measurements, and found to be largely consistent. 

The linear continua between 3.3$\,\mu$m and 3.6$\,\mu$m are indicated by dotted lines in Figure \ref{fig2}, from which the optical depth of the 3.4$\,\mu$m absorption feature was calculated. Since the optical depth is given by $\tau = -\ln$(I/I$_{0}$), where I is the measured flux and I$_{0}$ is the estimated continuum emission flux at 3.4\,$\mu$m, there is an uncertainty in optical depth values due to continuum flux estimations as we do not know the black body spectrum of the background stellar light source. This uncertainty has been mentioned in [I02], [M04] and [G12] and also discussed in detail in Paper 2. Moreover, the 3.4$\,\mu$m absorption feature is superimposed on the long wavelength wing of the broad 3.1$\,\mu$m  H$_{2}$O ice absorption band for lines of sight through Field A, B and C ([I02], [C02], [M04] and [G12]) and this adds another complication for the determination of the optical depth of 3.4\,$\mu$m absorption. To estimate \textit{the lowest limit} of aliphatic hydrocarbon absorptions, \textit{local linear continua} between 3.3$\,\mu$m to 3.6$\,\mu$m were applied to the spectra in [I02], [M04] and [G12]. Similarly, in our previous work, optical depths for Field A sources were calculated by interpolating across the aliphatic absorption feature from 3.3$\,\mu$m to 3.6$\,\mu$m to yield minimum aliphatic hydrocarbon absorptions. 

In this study we followed the same methodology for the optical depth measurements, so that we can compare our results with the measurements of Field A and the minimum levels of the aliphatic hydrocarbon absorptions reported in [I02] and [G12]. There are three sources in Field C: S7, S10 and S11, that have previously measured 3.4$\,\mu$m optical depths in [G12]. The brightest of these, S7, is partially saturated, therefore our derived values for it are not reliable. However, the results we obtained for the other two sources, S10 and S11, were found to be similar to the literature values given in [G12] (see Table~\ref{tab:3}). Then we compared our results with the optical depth measurements reported by [I02] for the line of sight through the IRAS 18511+0146 cluster. We found the measurements of S10 and S11 in agreement with the maximum level they reported despite the fact that their methods involve spectroscopic data (they measured aliphatic hydrocarbon absorptions around 3.4$\,\mu$m, produced by methylene and methyl groups separately) (see Table~\ref{tab:3}).

\begin{table*}
 \begin{center}
 \footnotesize
 \caption{Optical depth of the 3.4\,$\mu$m absorption feature for previously measured sources S7, S10 and S11 in Field C based on the GCIRS7--C02 calibration, compared to the values determined in [G12] and [I02].  } 
 \label{tab:3}
 \centering

 \begin{tabular}{| c | c | c | c |} 
\hline
\multirow{2}{*}{  } & \multicolumn{3}{c|}{$\tau_{3.4\,\mu m}$}  \\
\cline{2-4}
 & S7	& S10 & S11  \\
\hline
this study & 0.239 & 0.137 & 0.123 \\
\hline
[G12] & 0.073 & 0.093 & 0.119	\\
\hline
[I02] & \multicolumn{3}{c|}{ $0.087^{1}$ - $0.066^{2}$ }  \\
\hline

\end{tabular}
\end{center}
\begin{footnotesize}
The reported optical depths values for methylene (1) and methyl (2) groups, respectively. 
\end{footnotesize}
\end{table*}

We did, however, also check this calibration method with several other methods for determining the fluxes for the sources in Field C.  This included using the fluxes for sources S10 and S11 in the spectra of [G12] to provide the calibration (i.e.\ instead of using the spectrum of GCIRS7 from [C02]).  We also used the photometric fluxes for GCIRS7 in Field A and for S10 \& S11 in Field C,  obtained by interpolating the 2MASS K-band (2.17$\,\mu$m) and Spitzer IRAC\footnote{https://irsa.ipac.caltech.edu/data/SPITZER/} Ch1-band (3.6$\,\mu$m) fluxes \citep{Vig2007} onto the narrow filter bands for our observations (and including a correction for the flux at 3.4\,$\mu$m for the reported 3.4\,$\mu$m optical depths from [C02] and [G12]), to provide the flux calibration.  In all cases internally consistent results were found for each of these 5 additional calibration methods, with a near constant offset between the derived 3.4\,$\mu$m optical depths for all the sources in Field C found for each of these calibration methods and that determined when using fluxes derived for GCIRS7 from [C02];  these offsets range from 0.03 to 0.1 dependent on the particular calibration source chosen.  

These offsets are less than the differences in 3.4\,$\mu$m optical depths determined in Field A by different authors for the same source, as these authors apply different analysis methods to their data for a source.  For instance, for GCIRS7, [C02] and [C04] determined a 3.4\,$\mu$m optical depth of 0.15 and 0.41, respectively, using the different sets of spectral data they each obtained. \cite{Moultaka2004} also compared the derived 3.4\,$\mu$m optical depths for a given source applying different methods to estimate the continuum level of the spectrum. For example, for the source IRS16C they found the 3.4\,$\mu$m optical depths ranging from 0.14 to 0.49, depending on the method they used.

It is clear that it is difficult to obtain precise values for the optical depth, though the relative differences between sources are more reliably obtained, as we demonstrated in Paper 2\@.  Given these uncertainties, the offsets we obtained (i.e.\ from 0.03 to 0.1) are less than the variations in 3.4\,$\mu$m optical depth determined for the same source by the mentioned studies above.  Since we determined that using GCIRS7--C02 provided the best calibration set for Field A data in Paper 2, we have then applied it to the data for Fields B and C.

 \section{Mapping Applications}\label{Section4}
 
For mapping applications, 180 sources in Field B and for 15 sources in Field C (which are generally fainter than GC field sources) were selected to satisfy the S/N criteria.

The biases in aliphatic hydrocarbon maps arise primarily from two sources. The first is that the distances of the background sources are largely unknown since it is not possible to distinguish a reddened high mass, hot star from a low mass, cooler, intrinsically redder star. This means that we cannot be certain whether variations in the absorption are due to density or distances. The second bias arises because the data for each field are brightness limited. 

For the brightness limited bias, we split the data into quartiles according their brightness. We would like to note that there would be bias due to effective temperature and extinction degeneracy and some of the faintest objects could be the most reddened objects. However, we show below that the similar result was found independent of the brightness in the resultant quartile maps. 

The issue of where the sources lie was found to be not a serious concern since we could not find large source-to-source variations in maps prepared by using the quartile data sets. In the case of the Galactic Centre this is in fact the gas and dust in the central regions of the Galaxy. It is likely that most of the sources in Field A and Field B are at roughly the same distance and so have the same columns of absorbing material in front of them. For Field C, we have very limited number of sources to draw up a conclusion. However, since Field C samples IRAS 18511+0146 cluster sources, we can assume that most of the sources are sampling the same columns of absorbing material. 

We also compared the optical depths and the brightness of the sources by using the resultant maps. However, we could not find any correlation.

 \subsection{Application to a new field in the GC: Field B}

We present the fluxes for 180 sources measured at 3.3$\,\mu$m, 3.4$\,\mu$m and 3.6$\,\mu$m in Field B in Table \ref{tab:4} and the derived 3.4\,$\mu$m optical depths in Table \ref{tab:5}, together with their celestial coordinates. Optical depths were found to vary over a relatively small range across the field with the mean value of $0.36 \pm 0.09$.

We checked whether systematic biases due to flux levels of sources (to satisfy S/N=5) may have affected the optical depth determinations by dividing the sources into four quartiles based on their fluxes, similarly we applied previously for Field A (described in detail in Paper 2). The mean 3.4\,$\mu$m optical depths and standard deviation in each quartile are shown in Table \ref{tab:6}.  The differences between the derived 3.4\,$\mu$m optical depths are not significant, consistent with the flux level not biasing its determination.

\begin{table*}
 \begin{center}
  \caption{Calibrated fluxes ($\times$\,10$^{-18}$ W\,cm$^{-2}$ $\mu$m$^{-1}$) for the Field B sources used for 3.4$\,\mu$m optical depth calculations.} 
\centering
  \label{tab:4}
  \begin{tabular}{|  p {0.6 cm}  p {0.7 cm} p {0.7 cm} p {0.7 cm} | p {0.6 cm}  p {0.7 cm} p {0.7 cm} p {0.7 cm}  | p {0.6 cm}  p {0.7 cm} p {0.7 cm} p {0.7 cm} | p {0.6 cm}  p {0.7 cm} p {0.7 cm} p {0.7 cm} |}
    \hline

Source &  \multicolumn{3}{c|}{Fluxes } & Source & \multicolumn{3}{c|}{Fluxes } & Source & \multicolumn{3}{c|}{Fluxes}  &  Source &\multicolumn{3}{c|}{Fluxes} \\

\multirow{1}{*}{No} &	3.3$\,\mu$m & 	3.4$\,\mu$m 	&	3.6$\,\mu$m	&	\multirow{1}{*}{No} &	3.3$\,\mu$m & 	3.4$\,\mu$m	&3.6$\,\mu$m	 &\multirow{1}{*}{No} &	3.3$\,\mu$m & 	3.4$\,\mu$m 	&	3.6$\,\mu$m	& \multirow{1}{*}{No} &	3.3$\,\mu$m & 	3.4$\,\mu$m 	&	3.6$\,\mu$m	\\
    \hline

1	&	46.2	&	36.6	&	47.1	&	46	&	6.2	&	3.5	&	4.5	&	91	&	3.4	&	2.2	&	2.7	&	136	&	2.0	&	1.4	&	1.7	\\
2	&	48.4	&	35.6	&	42.4	&	47	&	4.8	&	3.5	&	4.4	&	92	&	4.3	&	2.0	&	2.7	&	137	&	2.0	&	1.3	&	1.6	\\
3	&	34.5	&	25.4	&	31.4	&	48	&	5.3	&	3.7	&	4.4	&	93	&	2.5	&	1.7	&	2.7	&	138	&	2.1	&	1.3	&	1.6	\\
4	&	30.2	&	16.6	&	24.2	&	49	&	5.1	&	3.6	&	4.4	&	94	&	3.3	&	2.3	&	2.7	&	139	&	2.0	&	1.5	&	1.6	\\
5	&	33.2	&	23.2	&	24.0	&	50	&	7.1	&	3.7	&	4.3	&	95	&	2.5	&	2.0	&	2.7	&	140	&	1.9	&	1.2	&	1.6	\\
6	&	18.9	&	14.2	&	21.1	&	51	&	4.4	&	3.2	&	4.3	&	96	&	3.2	&	2.2	&	2.7	&	141	&	1.7	&	1.1	&	1.6	\\
7	&	23.5	&	16.5	&	20.8	&	52	&	5.3	&	3.5	&	4.2	&	97	&	3.3	&	2.2	&	2.7	&	142	&	1.6	&	1.1	&	1.6	\\
8	&	20.4	&	14.0	&	20.4	&	53	&	3.8	&	2.6	&	4.1	&	98	&	2.0	&	1.4	&	2.6	&	143	&	1.6	&	1.0	&	1.6	\\
9	&	11.8	&	10.7	&	19.7	&	54	&	4.6	&	3.0	&	4.1	&	99	&	3.4	&	2.0	&	2.5	&	144	&	2.1	&	1.3	&	1.5	\\
10	&	15.0	&	11.3	&	16.7	&	55	&	4.6	&	2.9	&	4.1	&	100	&	3.0	&	1.7	&	2.5	&	145	&	1.8	&	1.2	&	1.5	\\
11	&	14.2	&	9.9	&	14.3	&	56	&	4.7	&	3.3	&	4.0	&	101	&	2.4	&	1.8	&	2.5	&	146	&	1.7	&	1.1	&	1.5	\\
12	&	21.5	&	12.2	&	13.4	&	57	&	3.7	&	2.8	&	4.0	&	102	&	2.7	&	1.9	&	2.5	&	147	&	1.9	&	1.2	&	1.5	\\
13	&	16.2	&	11.4	&	13.3	&	58	&	3.5	&	2.6	&	4.0	&	103	&	2.3	&	1.7	&	2.4	&	148	&	2.2	&	1.4	&	1.5	\\
14	&	11.9	&	8.5	&	12.1	&	59	&	4.7	&	3.0	&	4.0	&	104	&	2.5	&	1.6	&	2.4	&	149	&	1.3	&	0.9	&	1.5	\\
15	&	9.4	&	7.4	&	11.3	&	60	&	4.7	&	2.9	&	4.0	&	105	&	2.8	&	1.8	&	2.3	&	150	&	1.5	&	1.0	&	1.3	\\
16	&	12.0	&	7.8	&	10.9	&	61	&	4.6	&	2.8	&	3.9	&	106	&	3.1	&	1.6	&	2.3	&	151	&	1.6	&	1.2	&	1.3	\\
17	&	11.0	&	7.4	&	9.6	&	62	&	3.7	&	3.0	&	3.9	&	107	&	2.6	&	2.0	&	2.3	&	152	&	1.5	&	0.8	&	1.3	\\
18	&	12.3	&	6.8	&	9.4	&	63	&	4.9	&	2.9	&	3.9	&	108	&	2.5	&	1.6	&	2.3	&	153	&	1.9	&	1.0	&	1.3	\\
19	&	10.7	&	5.9	&	9.2	&	64	&	3.5	&	2.5	&	3.9	&	109	&	2.6	&	1.6	&	2.3	&	154	&	1.8	&	1.0	&	1.3	\\
20	&	9.9	&	6.3	&	8.7	&	65	&	4.3	&	2.8	&	3.9	&	110	&	2.6	&	1.8	&	2.3	&	155	&	1.6	&	1.2	&	1.3	\\
21	&	9.4	&	6.6	&	8.1	&	66	&	4.3	&	3.0	&	3.9	&	111	&	2.8	&	1.8	&	2.2	&	156	&	1.6	&	0.9	&	1.3	\\
22	&	9.8	&	5.9	&	8.1	&	67	&	3.0	&	2.5	&	3.8	&	112	&	2.8	&	1.9	&	2.2	&	157	&	1.7	&	1.0	&	1.2	\\
23	&	9.8	&	6.4	&	7.8	&	68	&	4.4	&	3.1	&	3.8	&	113	&	2.9	&	1.8	&	2.2	&	158	&	1.4	&	1.0	&	1.2	\\
24	&	9.1	&	6.5	&	7.7	&	69	&	4.1	&	2.8	&	3.7	&	114	&	3.0	&	1.6	&	2.2	&	159	&	1.3	&	0.9	&	1.1	\\
25	&	7.1	&	5.1	&	6.7	&	70	&	4.1	&	3.0	&	3.5	&	115	&	2.7	&	1.8	&	2.2	&	160	&	1.2	&	0.8	&	1.1	\\
26	&	8.3	&	5.5	&	6.5	&	71	&	3.9	&	2.7	&	3.5	&	116	&	2.0	&	1.5	&	2.2	&	161	&	1.5	&	0.8	&	1.1	\\
27	&	7.8	&	5.6	&	6.5	&	72	&	4.2	&	2.8	&	3.5	&	117	&	2.2	&	1.5	&	2.1	&	162	&	1.4	&	0.7	&	1.1	\\
28	&	6.8	&	4.9	&	6.5	&	73	&	4.5	&	3.0	&	3.5	&	118	&	2.3	&	1.5	&	2.1	&	163	&	1.6	&	0.8	&	1.1	\\
29	&	7.9	&	4.8	&	6.4	&	74	&	3.7	&	2.6	&	3.3	&	119	&	2.6	&	1.7	&	2.1	&	164	&	1.5	&	0.9	&	1.0	\\
30	&	5.6	&	4.5	&	6.3	&	75	&	4.0	&	2.8	&	3.3	&	120	&	2.7	&	1.6	&	2.1	&	165	&	1.3	&	0.9	&	1.0	\\
31	&	7.6	&	5.0	&	6.2	&	76	&	4.0	&	2.6	&	3.2	&	121	&	2.4	&	1.5	&	2.1	&	166	&	1.1	&	0.8	&	1.0	\\
32	&	6.0	&	4.4	&	6.2	&	77	&	3.7	&	2.3	&	3.2	&	122	&	2.8	&	1.7	&	2.0	&	167	&	1.5	&	0.9	&	1.0	\\
33	&	4.4	&	3.7	&	6.1	&	78	&	3.9	&	2.5	&	3.1	&	123	&	2.3	&	1.5	&	2.0	&	168	&	1.3	&	0.6	&	1.0	\\
34	&	6.5	&	4.5	&	5.8	&	79	&	3.5	&	2.5	&	3.1	&	124	&	2.4	&	1.6	&	1.9	&	169	&	1.2	&	0.7	&	1.0	\\
35	&	6.9	&	4.4	&	5.8	&	80	&	3.8	&	2.2	&	3.0	&	125	&	2.5	&	1.6	&	1.9	&	170	&	1.1	&	0.7	&	1.0	\\
36	&	5.5	&	4.1	&	5.5	&	81	&	3.5	&	2.5	&	3.0	&	126	&	2.5	&	1.4	&	1.9	&	171	&	1.3	&	0.8	&	1.0	\\
37	&	7.2	&	3.7	&	5.4	&	82	&	2.8	&	1.9	&	3.0	&	127	&	2.0	&	1.3	&	1.8	&	172	&	1.1	&	0.8	&	1.0	\\
38	&	7.6	&	5.3	&	5.1	&	83	&	3.5	&	2.3	&	3.0	&	128	&	2.2	&	1.4	&	1.8	&	173	&	1.3	&	0.8	&	1.0	\\
39	&	4.7	&	3.3	&	5.1	&	84	&	3.6	&	2.5	&	3.0	&	129	&	2.3	&	1.6	&	1.8	&	174	&	1.1	&	0.7	&	0.9	\\
40	&	5.6	&	3.9	&	5.0	&	85	&	4.4	&	2.1	&	2.9	&	130	&	2.2	&	1.5	&	1.8	&	175	&	1.2	&	0.6	&	0.9	\\
41	&	5.8	&	4.2	&	4.9	&	86	&	3.7	&	2.2	&	2.9	&	131	&	2.0	&	1.3	&	1.8	&	176	&	1.0	&	0.6	&	0.9	\\
42	&	5.5	&	3.4	&	4.9	&	87	&	3.4	&	2.3	&	2.8	&	132	&	2.0	&	1.3	&	1.7	&	177	&	0.9	&	0.6	&	0.9	\\
43	&	4.9	&	3.6	&	4.7	&	88	&	3.6	&	2.1	&	2.8	&	133	&	2.1	&	1.4	&	1.7	&	178	&	1.1	&	0.7	&	0.8	\\
44	&	5.7	&	3.8	&	4.7	&	89	&	3.3	&	2.2	&	2.8	&	134	&	2.0	&	1.3	&	1.7	&	179	&	0.7	&	0.6	&	0.8	\\
45	&	4.1	&	2.8	&	4.5	&	90	&	3.2	&	2.3	&	2.7	&	135	&	1.9	&	1.4	&	1.7	&	180	&	1.0	&	0.6	&	0.8	\\

\hline
\end{tabular}
    \end{center}
  \end{table*}

\begin{table*}
 \begin{center}
  \caption{Celestial coordinates (J2000) in degrees and the optical depths for the 3.4$\,\mu$m absorption feature for the Field B sources.} 
\centering
  \label{tab:5}
   \begin{tabular}{| p {0.3 cm}  p {0.9 cm} p {1.1 cm} p {0.4 cm} | p {0.3 cm}  p {0.9 cm} p {1.1 cm} p {0.4 cm}  | p {0.3 cm}  p {0.9 cm} p {1.1 cm} p {0.4 cm}  | p {0.3 cm}  p {0.9 cm} p {1.1 cm} p {0.4 cm} |}
    \hline

No & 	RA & Dec	&	$\tau_{3.4}$	&	No  &	RA & Dec	&	$\tau_{3.4}$	&  No &	RA & Dec	&	$\tau_{3.4}$	&  No &	RA & Dec 	&	$\tau_{3.4}$	\\
    \hline

1	&	266.3042	&	-29.1602	&	0.18	&	46	&	266.2836	&	-29.1629	&	0.40	&	91	&	266.2875	&	-29.1705	&	0.30	&	136	&	266.3030	&	-29.1611	&	0.24	\\
2	&	266.3025	&	-29.1626	&	0.21	&	47	&	266.3126	&	-29.1699	&	0.22	&	92	&	266.2832	&	-29.1448	&	0.53	&	137	&	266.3142	&	-29.1482	&	0.34	\\
3	&	266.3159	&	-29.1596	&	0.22	&	48	&	266.3279	&	-29.1647	&	0.24	&	93	&	266.2914	&	-29.1578	&	0.34	&	138	&	266.2883	&	-29.1472	&	0.36	\\
4	&	266.2910	&	-29.1410	&	0.46	&	49	&	266.3039	&	-29.1798	&	0.22	&	94	&	266.3055	&	-29.1532	&	0.27	&	139	&	266.3082	&	-29.1708	&	0.19	\\
5	&	266.3010	&	-29.1593	&	0.21	&	50	&	266.2876	&	-29.1405	&	0.47	&	95	&	266.3076	&	-29.1646	&	0.21	&	140	&	266.3146	&	-29.1802	&	0.31	\\
6	&	266.3240	&	-29.1552	&	0.27	&	51	&	266.3311	&	-29.1623	&	0.24	&	96	&	266.3011	&	-29.1587	&	0.27	&	141	&	266.2961	&	-29.1492	&	0.36	\\
7	&	266.3170	&	-29.1558	&	0.25	&	52	&	266.3264	&	-29.1743	&	0.25	&	97	&	266.3084	&	-29.1496	&	0.31	&	142	&	266.3118	&	-29.1658	&	0.27	\\
8	&	266.2954	&	-29.1552	&	0.32	&	53	&	266.3081	&	-29.1404	&	0.35	&	98	&	266.3254	&	-29.1505	&	0.40	&	143	&	266.2952	&	-29.1511	&	0.37	\\
9	&	266.3098	&	-29.1553	&	0.24	&	54	&	266.3212	&	-29.1406	&	0.32	&	99	&	266.2858	&	-29.1595	&	0.39	&	144	&	266.3264	&	-29.1473	&	0.30	\\
10	&	266.2942	&	-29.1709	&	0.25	&	55	&	266.3158	&	-29.1403	&	0.37	&	100	&	266.2883	&	-29.1442	&	0.46	&	145	&	266.2868	&	-29.1729	&	0.31	\\
11	&	266.2932	&	-29.1643	&	0.30	&	56	&	266.3254	&	-29.1623	&	0.24	&	101	&	266.3112	&	-29.1722	&	0.23	&	146	&	266.3279	&	-29.1550	&	0.33	\\
12	&	266.2844	&	-29.1589	&	0.36	&	57	&	266.3032	&	-29.1795	&	0.25	&	102	&	266.3022	&	-29.1558	&	0.27	&	147	&	266.3191	&	-29.1600	&	0.27	\\
13	&	266.3212	&	-29.1770	&	0.23	&	58	&	266.2929	&	-29.1760	&	0.26	&	103	&	266.3063	&	-29.1772	&	0.27	&	148	&	266.3210	&	-29.1479	&	0.27	\\
14	&	266.3246	&	-29.1454	&	0.28	&	59	&	266.2985	&	-29.1775	&	0.30	&	104	&	266.3285	&	-29.1793	&	0.30	&	149	&	266.3251	&	-29.1395	&	0.38	\\
15	&	266.2986	&	-29.1668	&	0.24	&	60	&	266.3311	&	-29.1443	&	0.35	&	105	&	266.2853	&	-29.1564	&	0.33	&	150	&	266.3228	&	-29.1427	&	0.34	\\
16	&	266.2872	&	-29.1621	&	0.33	&	61	&	266.2830	&	-29.1721	&	0.34	&	106	&	266.2912	&	-29.1421	&	0.52	&	151	&	266.3232	&	-29.1639	&	0.22	\\
17	&	266.3064	&	-29.1449	&	0.30	&	62	&	266.3107	&	-29.1608	&	0.19	&	107	&	266.3211	&	-29.1619	&	0.18	&	152	&	266.3223	&	-29.1381	&	0.45	\\
18	&	266.2898	&	-29.1459	&	0.44	&	63	&	266.3148	&	-29.1377	&	0.39	&	108	&	266.2943	&	-29.1548	&	0.35	&	153	&	266.2914	&	-29.1437	&	0.48	\\
19	&	266.2885	&	-29.1381	&	0.50	&	64	&	266.3140	&	-29.1455	&	0.33	&	109	&	266.2988	&	-29.1491	&	0.37	&	154	&	266.2831	&	-29.1653	&	0.39	\\
20	&	266.3188	&	-29.1436	&	0.34	&	65	&	266.3091	&	-29.1432	&	0.33	&	110	&	266.3020	&	-29.1699	&	0.25	&	155	&	266.3210	&	-29.1556	&	0.18	\\
21	&	266.3223	&	-29.1720	&	0.24	&	66	&	266.3152	&	-29.1674	&	0.25	&	111	&	266.3074	&	-29.1513	&	0.30	&	156	&	266.3022	&	-29.1437	&	0.38	\\
22	&	266.2829	&	-29.1776	&	0.33	&	67	&	266.3049	&	-29.1607	&	0.20	&	112	&	266.2988	&	-29.1479	&	0.27	&	157	&	266.2828	&	-29.1594	&	0.40	\\
23	&	266.3239	&	-29.1513	&	0.29	&	68	&	266.3147	&	-29.1582	&	0.24	&	113	&	266.2876	&	-29.1549	&	0.33	&	158	&	266.2995	&	-29.1514	&	0.25	\\
24	&	266.3206	&	-29.1682	&	0.22	&	69	&	266.3218	&	-29.1522	&	0.28	&	114	&	266.2846	&	-29.1485	&	0.42	&	159	&	266.3224	&	-29.1787	&	0.24	\\
25	&	266.3170	&	-29.1524	&	0.25	&	70	&	266.3124	&	-29.1790	&	0.21	&	115	&	266.3180	&	-29.1782	&	0.26	&	160	&	266.3192	&	-29.1482	&	0.31	\\
26	&	266.3125	&	-29.1510	&	0.29	&	71	&	266.3013	&	-29.1668	&	0.27	&	116	&	266.3299	&	-29.1723	&	0.27	&	161	&	266.2973	&	-29.1782	&	0.44	\\
27	&	266.3134	&	-29.1536	&	0.22	&	72	&	266.3201	&	-29.1461	&	0.28	&	117	&	266.2953	&	-29.1683	&	0.29	&	162	&	266.2874	&	-29.1602	&	0.40	\\
28	&	266.3086	&	-29.1775	&	0.25	&	73	&	266.3181	&	-29.1712	&	0.26	&	118	&	266.2922	&	-29.1608	&	0.31	&	163	&	266.2993	&	-29.1373	&	0.47	\\
29	&	266.2853	&	-29.1736	&	0.33	&	74	&	266.3234	&	-29.1597	&	0.26	&	119	&	266.3201	&	-29.1513	&	0.34	&	164	&	266.2963	&	-29.1803	&	0.34	\\
30	&	266.3129	&	-29.1626	&	0.20	&	75	&	266.3079	&	-29.1555	&	0.24	&	120	&	266.3297	&	-29.1790	&	0.37	&	165	&	266.2990	&	-29.1569	&	0.26	\\
31	&	266.3158	&	-29.1466	&	0.30	&	76	&	266.3171	&	-29.1639	&	0.31	&	121	&	266.2967	&	-29.1581	&	0.36	&	166	&	266.2988	&	-29.1639	&	0.22	\\
32	&	266.3286	&	-29.1543	&	0.24	&	77	&	266.2890	&	-29.1669	&	0.34	&	122	&	266.2921	&	-29.1736	&	0.34	&	167	&	266.3283	&	-29.1702	&	0.28	\\
33	&	266.2943	&	-29.1658	&	0.21	&	78	&	266.3247	&	-29.1672	&	0.28	&	123	&	266.2965	&	-29.1784	&	0.29	&	168	&	266.2843	&	-29.1427	&	0.47	\\
34	&	266.3211	&	-29.1594	&	0.26	&	79	&	266.3214	&	-29.1570	&	0.25	&	124	&	266.3089	&	-29.1504	&	0.29	&	169	&	266.2896	&	-29.1451	&	0.56	\\
35	&	266.3309	&	-29.1767	&	0.32	&	80	&	266.2880	&	-29.1508	&	0.40	&	125	&	266.3212	&	-29.1444	&	0.28	&	170	&	266.3102	&	-29.1664	&	0.32	\\
36	&	266.3089	&	-29.1791	&	0.24	&	81	&	266.3080	&	-29.1546	&	0.23	&	126	&	266.2963	&	-29.1390	&	0.44	&	171	&	266.3095	&	-29.1575	&	0.28	\\
37	&	266.2922	&	-29.1408	&	0.50	&	82	&	266.2913	&	-29.1525	&	0.37	&	127	&	266.2937	&	-29.1622	&	0.36	&	172	&	266.2973	&	-29.1737	&	0.24	\\
38	&	266.3149	&	-29.1574	&	0.19	&	83	&	266.3162	&	-29.1496	&	0.30	&	128	&	266.2928	&	-29.1781	&	0.30	&	173	&	266.3036	&	-29.1701	&	0.30	\\
39	&	266.3289	&	-29.1461	&	0.30	&	84	&	266.3205	&	-29.1546	&	0.28	&	129	&	266.3272	&	-29.1751	&	0.24	&	174	&	266.2953	&	-29.1531	&	0.33	\\
40	&	266.3229	&	-29.1518	&	0.26	&	85	&	266.2908	&	-29.1378	&	0.54	&	130	&	266.2871	&	-29.1779	&	0.27	&	175	&	266.2845	&	-29.1509	&	0.40	\\
41	&	266.3135	&	-29.1565	&	0.22	&	86	&	266.3138	&	-29.1428	&	0.37	&	131	&	266.3267	&	-29.1570	&	0.33	&	176	&	266.2906	&	-29.1760	&	0.36	\\
42	&	266.2846	&	-29.1626	&	0.38	&	87	&	266.3177	&	-29.1619	&	0.26	&	132	&	266.2986	&	-29.1597	&	0.32	&	177	&	266.2994	&	-29.1678	&	0.24	\\
43	&	266.3255	&	-29.1609	&	0.24	&	88	&	266.2903	&	-29.1673	&	0.35	&	133	&	266.2988	&	-29.1469	&	0.27	&	178	&	266.3222	&	-29.1605	&	0.24	\\
44	&	266.3165	&	-29.1542	&	0.29	&	89	&	266.3082	&	-29.1451	&	0.28	&	134	&	266.2957	&	-29.1596	&	0.32	&	179	&	266.3021	&	-29.1534	&	0.26	\\
45	&	266.2933	&	-29.1503	&	0.37	&	90	&	266.3228	&	-29.1607	&	0.26	&	135	&	266.3003	&	-29.1667	&	0.20	&	180	&	266.3208	&	-29.1728	&	0.37	\\	
																																																									
\hline
\end{tabular}
    \end{center}
  \end{table*}

 \begin{table*}
 \begin{center}
 \footnotesize
  \caption{Mean fluxes ($\times$\,10$^{-18}$ W\,cm$^{-2}$ $\mu$m$^{-1}$) and 3.4 $\mu$m optical depths ($\tau_{3.4}$) for each of the quartiles in Field B.}
  \label{tab:6}
  \centering 
 \begin{tabular}{| c | c | c | c | c | }
     \hline 
    & Flux Group 1 &   Flux Group 2 &   Flux Group 3 &   Flux Group 4 \\
  \hline 
  Flux  & 	11.2	&	3.5	&	2.2	&	1.2	\\	
  \hline 
$\tau_{3.4\,\mu m}$ & 0.35$\pm$0.08  & 0.36 $\pm$0.07 & 0.39 $\pm$0.07  & 0.40$\pm$0.10	\\

\hline
\end{tabular}
\end{center}
\end{table*}

\begin{figure*}
  \begin{center}
    \begin{tabular}{cc}
      {\includegraphics[angle=0,scale=0.42]{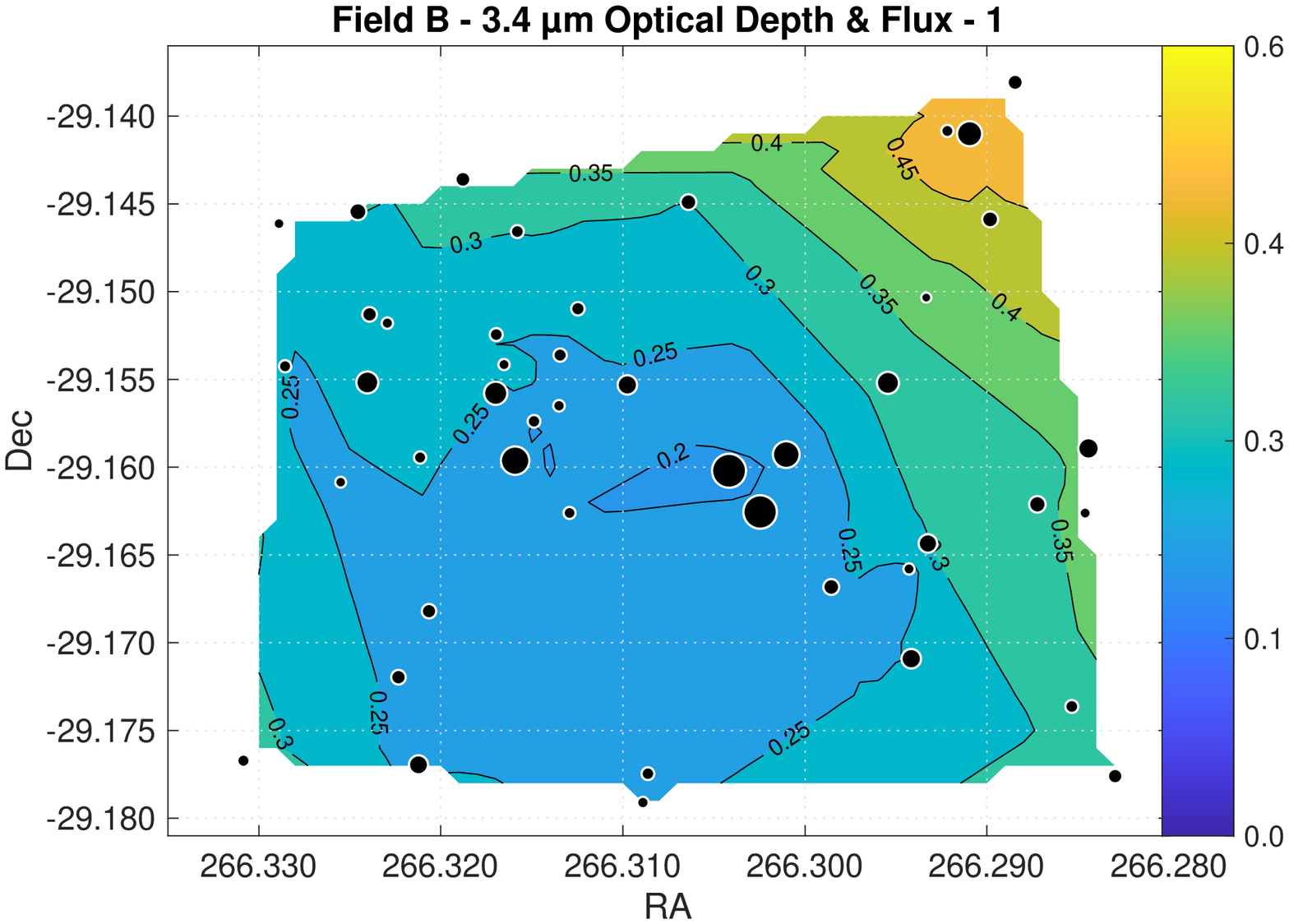}} &
      {\includegraphics[angle=0,scale=0.42]{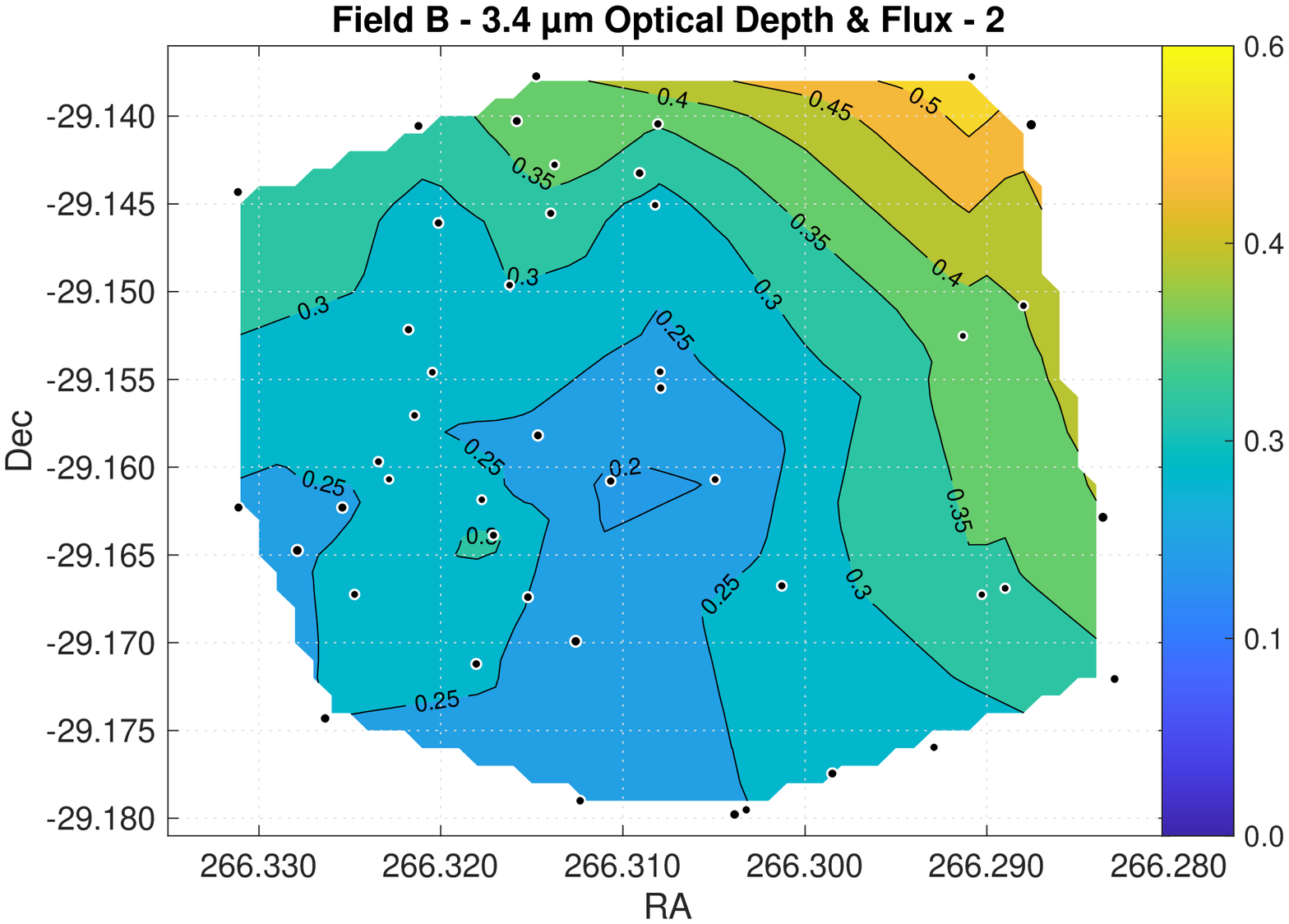}} \\
      {\includegraphics[angle=0,scale=0.42]{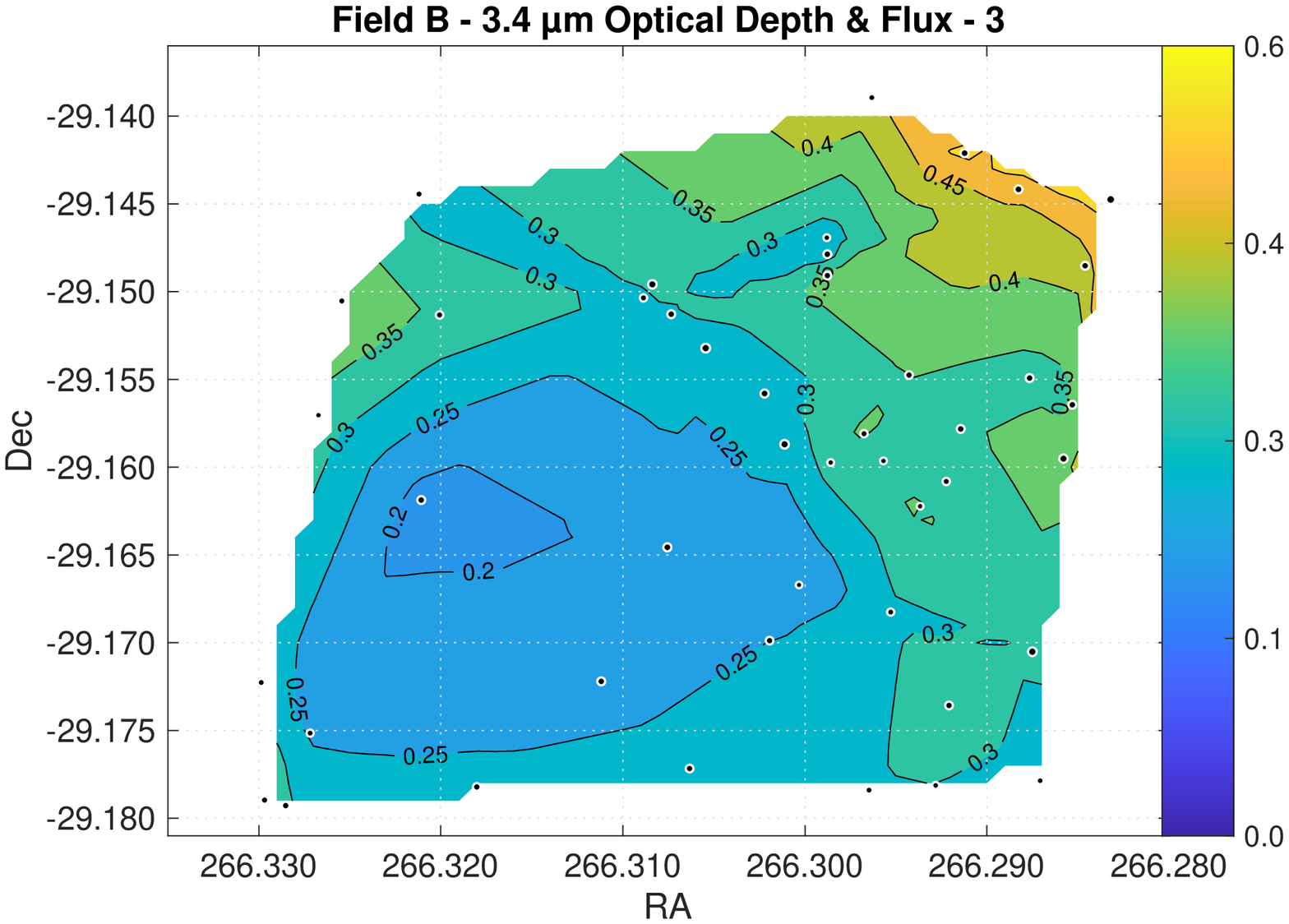}} &
      {\includegraphics[angle=0,scale=0.42]{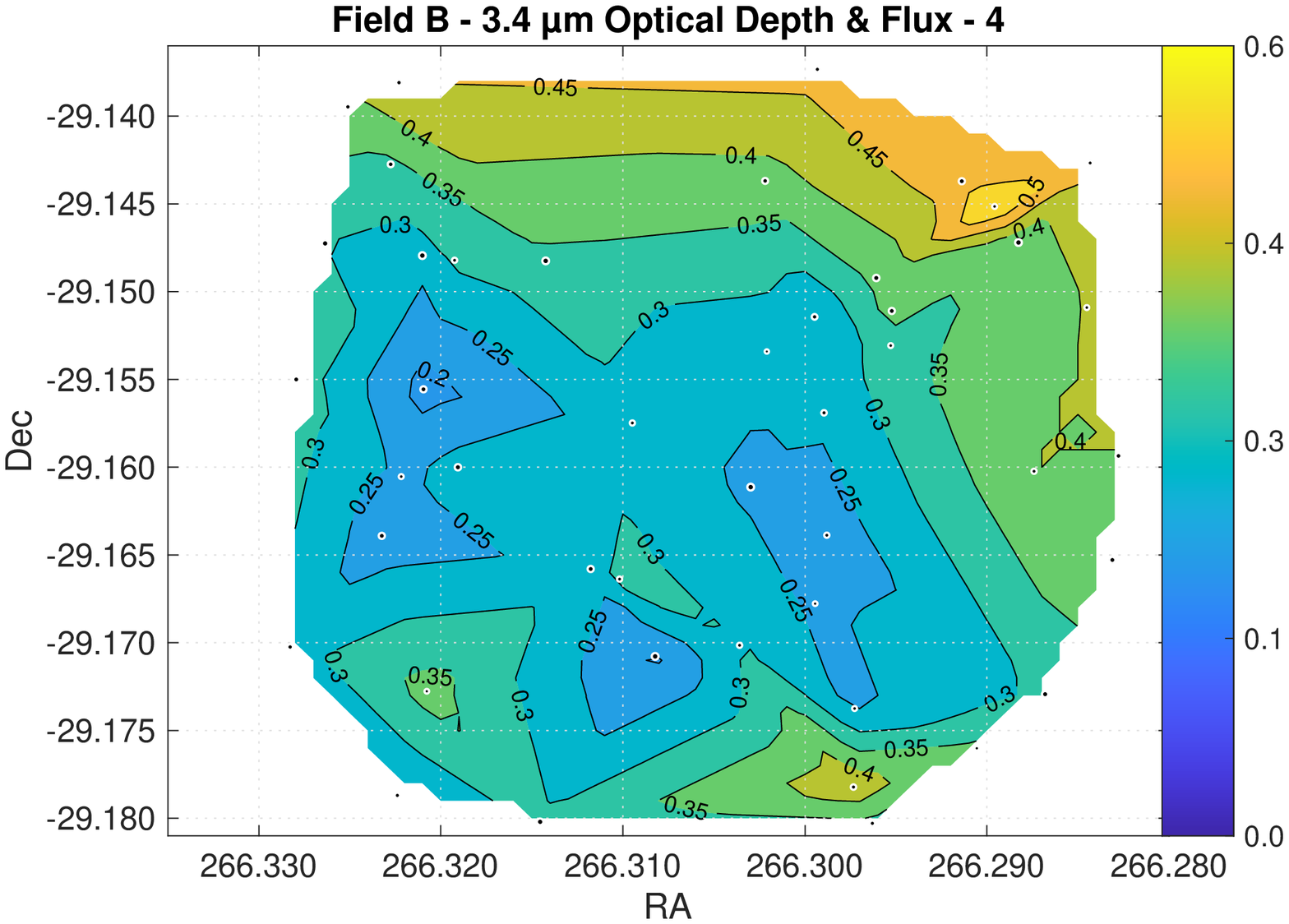}} \\
    \end{tabular}
    \caption{Map of the optical depth from 3.4$\,\mu$m absorption feature for the sources in the four flux quartiles in Field B, from brightest (group 1) to faintest (group 4).  Each map contains 45 sources.  Colour bars and contours indicate the $\tau_{3.4\,\mu m}$ levels. The location of sources are shown by black dots whose sizes are proportional to their fluxes.}   
   
    \label{fig3}
      \end{center}
\end{figure*}

We then considered the spatial distribution of the 3.4\,$\mu$m optical depth for the 45 sources in each flux (at 3.6\,$\mu$m) quartile range within Field B, with maps shown in Figure~\ref{fig3}. In this way, the effect of any possible individual erroneous measurement was also examined. While there is clearly some scatter in the distribution as there are areas within the field with few sources, the same pattern is apparent in each of the quartile ranges. 

There is a gradient across the field from SE to NW, with the 3.4\,$\mu$m optical depth rising approximately 50\% from one side to the other. This is independent of the brightness of individual sources and thus not likely to arise from the larger uncertainties in 3.4\,$\mu$m optical depth derived from the fainter sources. The faintest flux group does indeed show more variation in the pattern, most likely due to lower S/N in this group.

\begin{figure*}
  \begin{center}
    \begin{tabular}{cc}
      {\includegraphics[angle=0,scale=0.42]{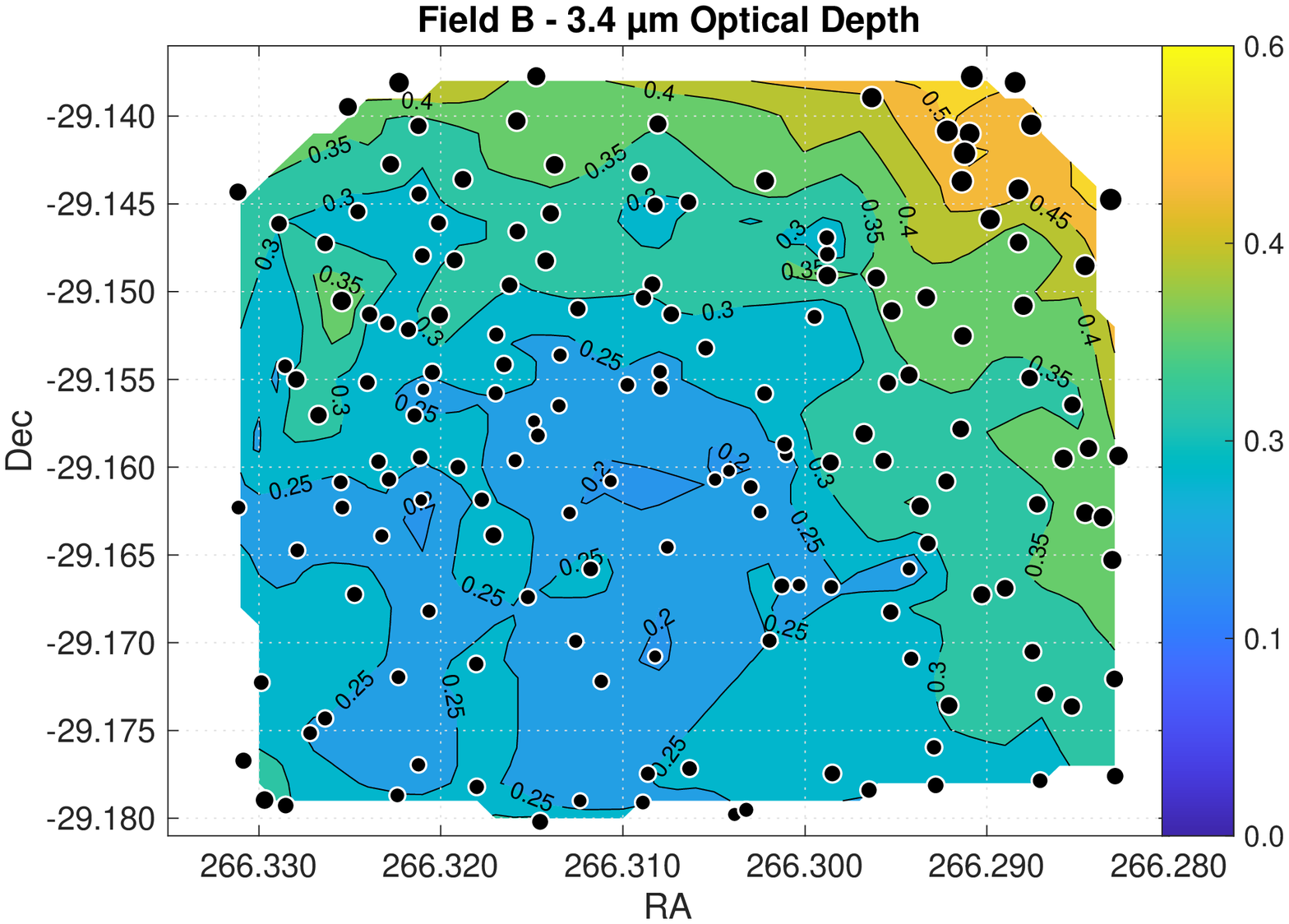}} &
      {\includegraphics[angle=0,scale=0.42]{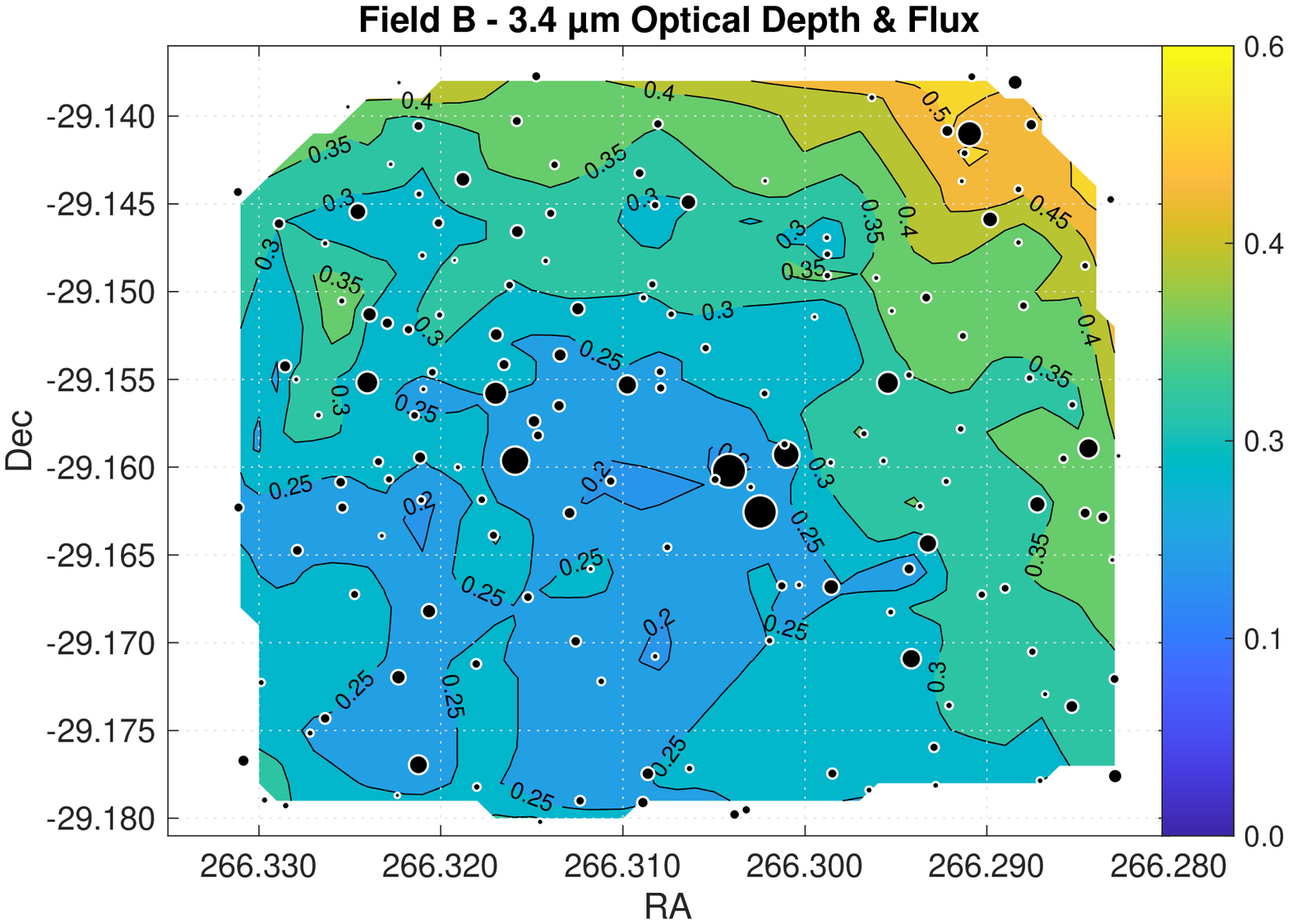}} \\
        \end{tabular}
    \caption{Map of the optical depth in 3.4$\,\mu$m absorption feature for Field B in celestial coordinates (J2000) for the brightest 160 sources. Colour bars and contours indicate $\tau_{3.4\,\mu m}$ levels. On the left panel location of the sources are shown by black dots whose sizes are proportional to the optical depth values. On the right panel, sizes are proportional to the 3.4$\,\mu$m flux values.}   
    \label{fig4}
      \end{center}
\end{figure*}

We conclude that we are able to reliably measure the 3.4\,$\mu$m optical depth using this technique of imaging through narrow band filters in Field B, to a sensitivity level of $\sim$10$^{-18}$ W\,cm$^{-2}$ $\mu$m$^{-1}$ at 3.6$\,\mu$m ($\sim$10 mag). Given the increased scatter in the lowest flux quartile we remove the faintest 20 sources from the list to take the brightest 160 sources and plot their 3.4\,$\mu$m optical depths in Figure \ref{fig4}. The location of the background sources are shown by black dots. On the left panel size of the dots are proportional to their 3.4\,$\mu$m optical depth in the line of sights. On the right panel size of the dots are proportional to the flux (at 3.6\,$\mu$m) of the sources. The optical depth, in general, is seen to rise from $\sim 0.2$ to $\sim 0.6$ moving from SE to NW across this field near the centre of the Galaxy.

There could readily be local source-to-source variations in 3.4\,$\mu$m optical depth, superimposed on a broader trend. However, examination of Figure \ref{fig4} (right panel), does not suggest this to be significant as little inter-source scatter is apparent.

\subsection{Application to a field in the Galactic Plane: Field C (IRAS 18511+0146)}
 
\begin{figure*}
  \begin{center}
    \begin{tabular}{cc}
      {\includegraphics[angle=0,scale=0.42]{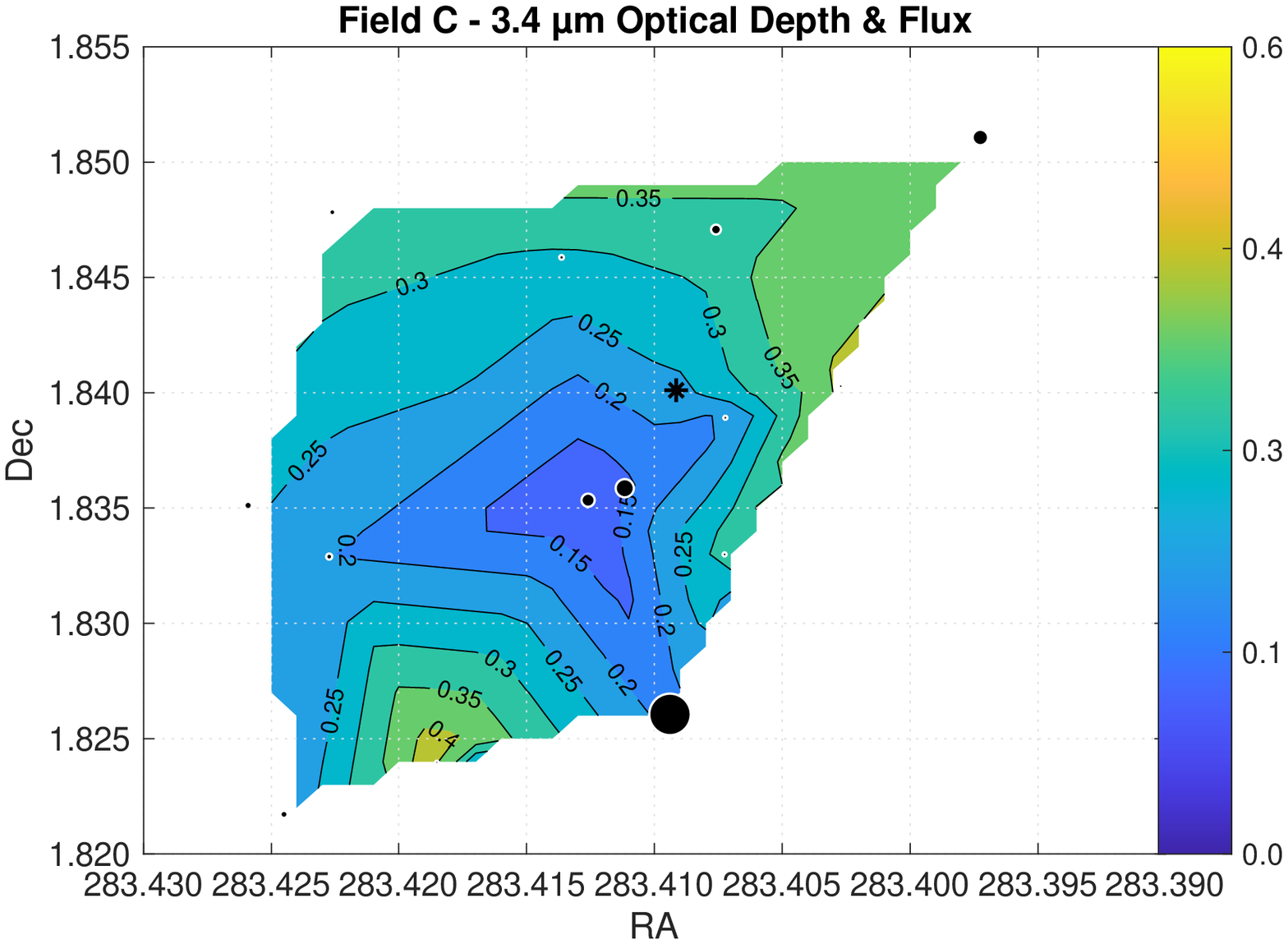}} &
      {\includegraphics[angle=0,scale=0.42]{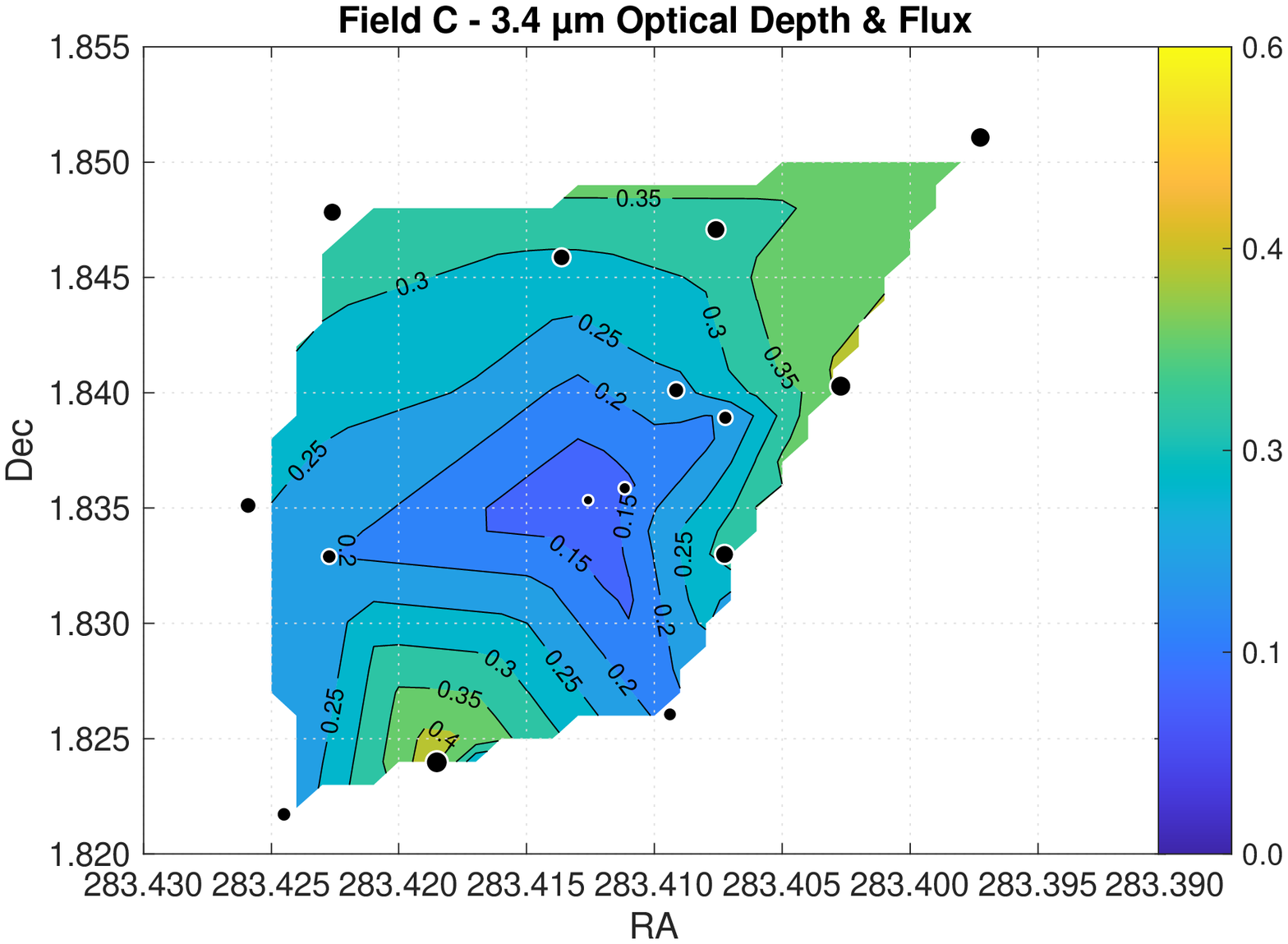}} \\
        \end{tabular}
    \caption{Map of the optical depth from 3.4$\,\mu$m absorption feature for the Field C in celestial coordinates (J2000). Colour bars and contours indicate the $\tau_{3.4\,\mu m}$ levels. On the left panel locations of sources are shown by black dots whose sizes are proportional to the flux values (we note that the brightness of S7 is not shown but its location indicated by a star ($\ast$) as it is far brighter than the other sources and partly saturated). On the right panel instead, these sizes are proportional to the 3.4$\,\mu$m optical depth.}   
    \label{fig5}
      \end{center}
\end{figure*}

We used the resultant spectra of 15 sources to provide a low resolution map of the 3.4$\,\mu$m optical depth across the Field C in Figure \ref{fig5}. Colour bars and contours indicate the 3.4\,$\mu$m optical depth levels, which were found to range from 0.1 to 0.4 across the field.  On the left panel, the location of the background sources are shown and size of the dots are proportional to the flux (at 3.6\,$\mu$m) of the sources (the brightness of S7 is not shown but its location indicated by a \textit{star symbol} ($\ast$) as it is far brighter than the other sources and partly saturated). On the right panel, size of the dots are proportional to their 3.4$\,\mu$m optical depth in the line of sight. There is no correlation between fluxes and optical depths, indicating no obvious bias in the determination of the latter, though we note that the care must be taken in interpretation given the low number of irregularly spaced points (15) used in the map's creation.

For the Galactic Centre region, maps obtained with four different sets of stars are compared to each other and found to be consistent. However, we could not apply this test to Field C as there is an insufficient number of sources.

	\begin{figure*}
  \begin{center}
    \begin{tabular}{cc}
      {\includegraphics[angle=0,scale=0.46]{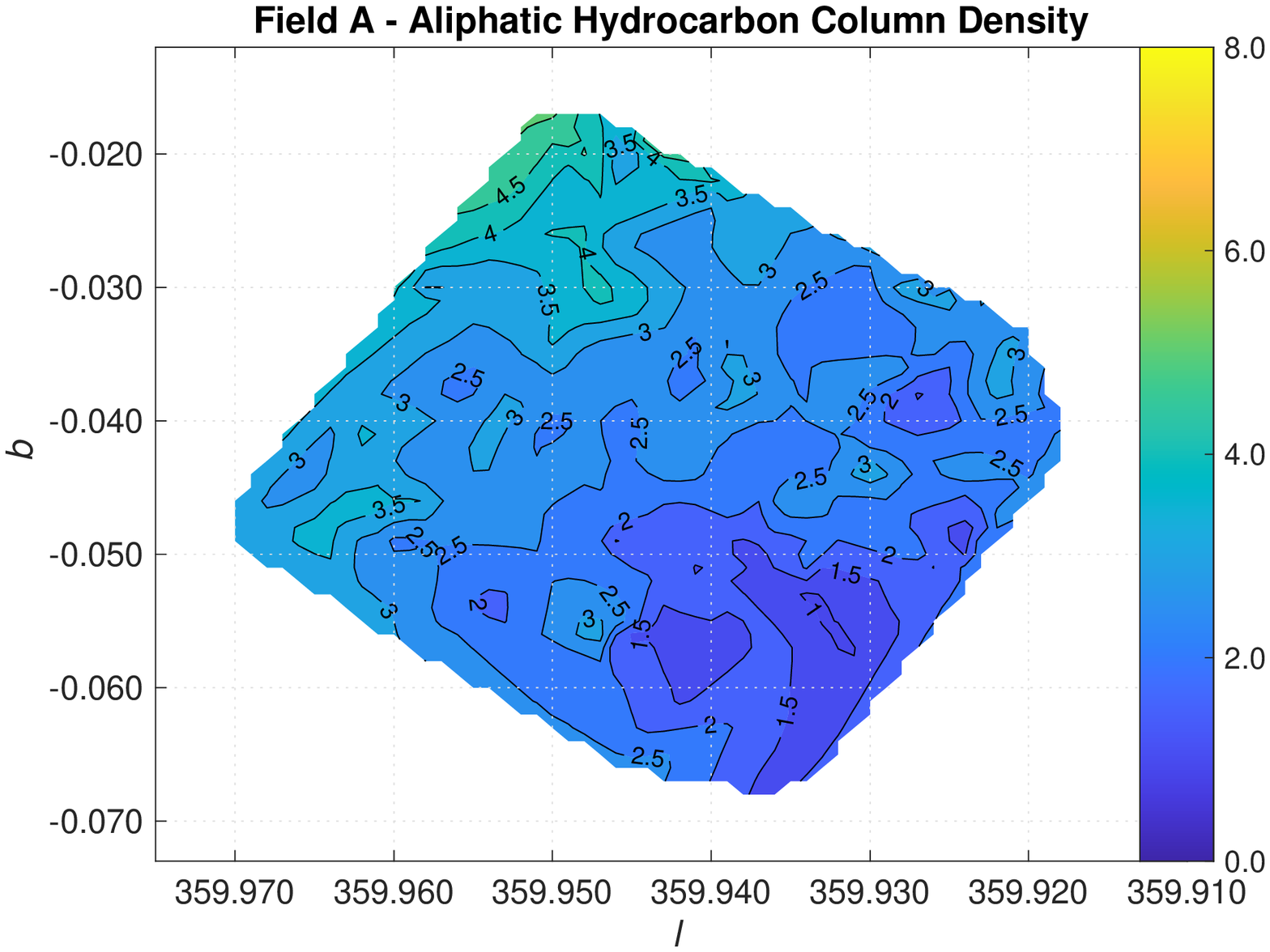}} & {\includegraphics[angle=0,scale=0.46]{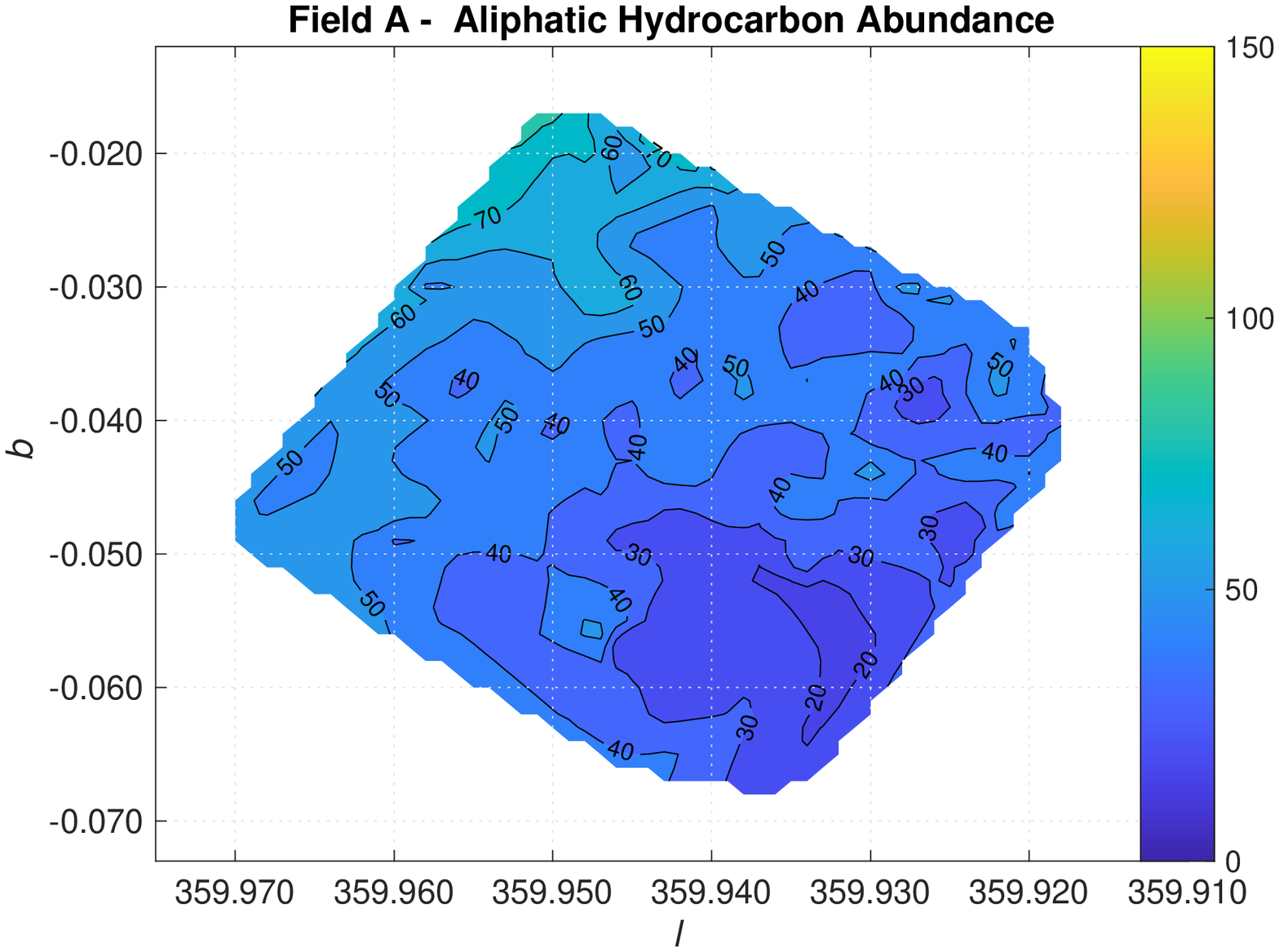}} \\
        {\includegraphics[angle=0,scale=0.46]{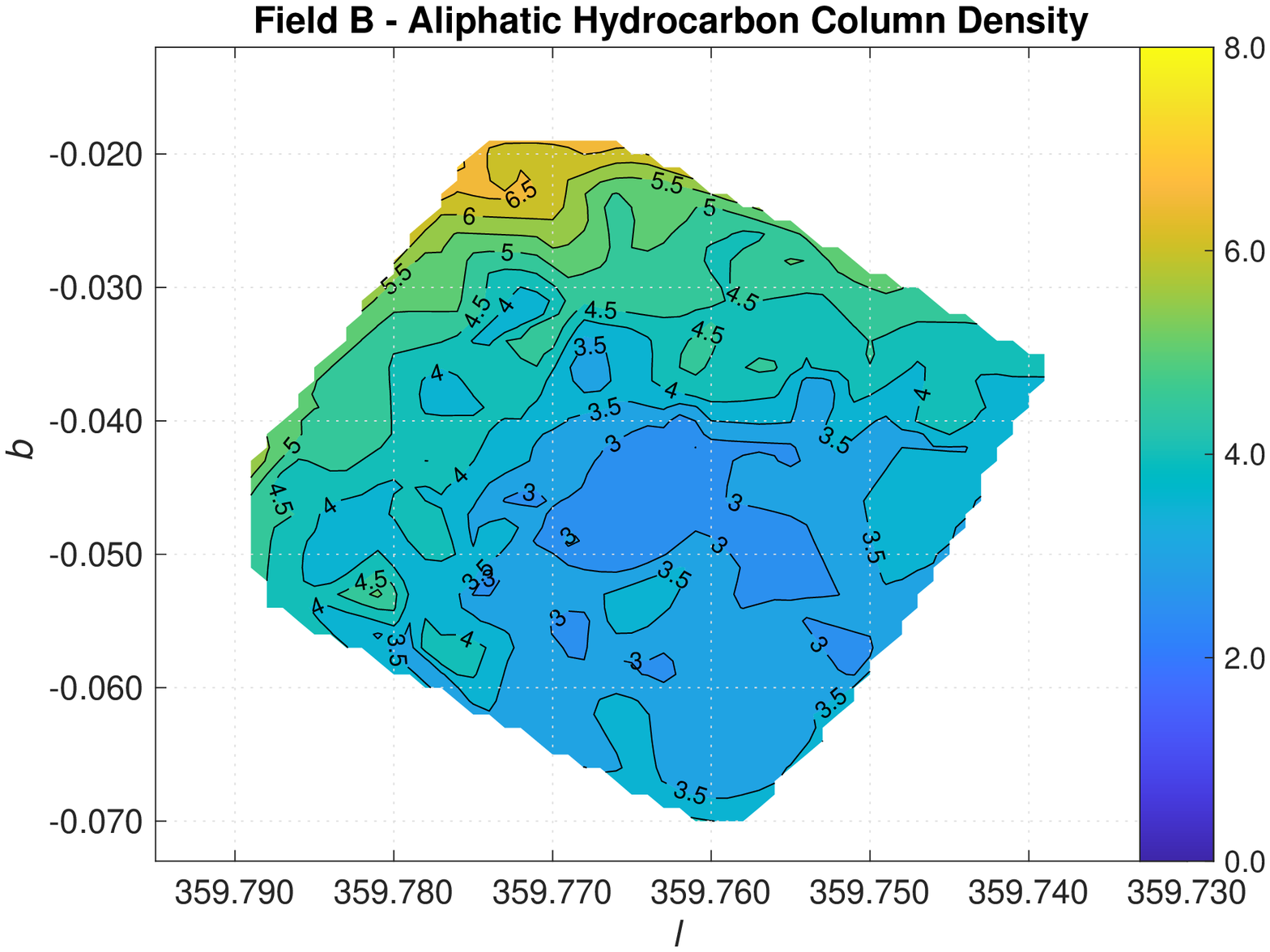}} & {\includegraphics[angle=0,scale=0.46]{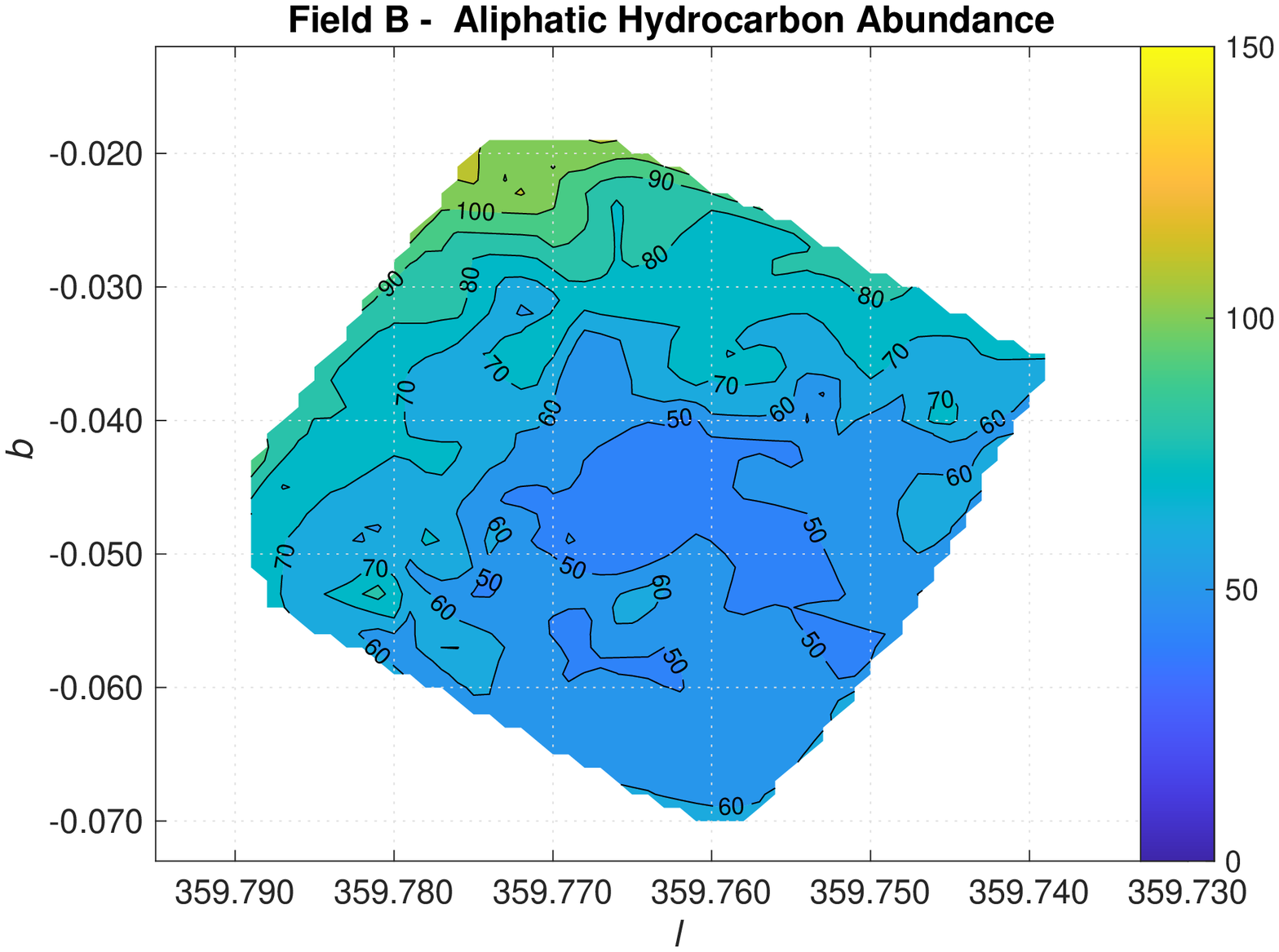}} \\
          {\includegraphics[angle=0,scale=0.46]{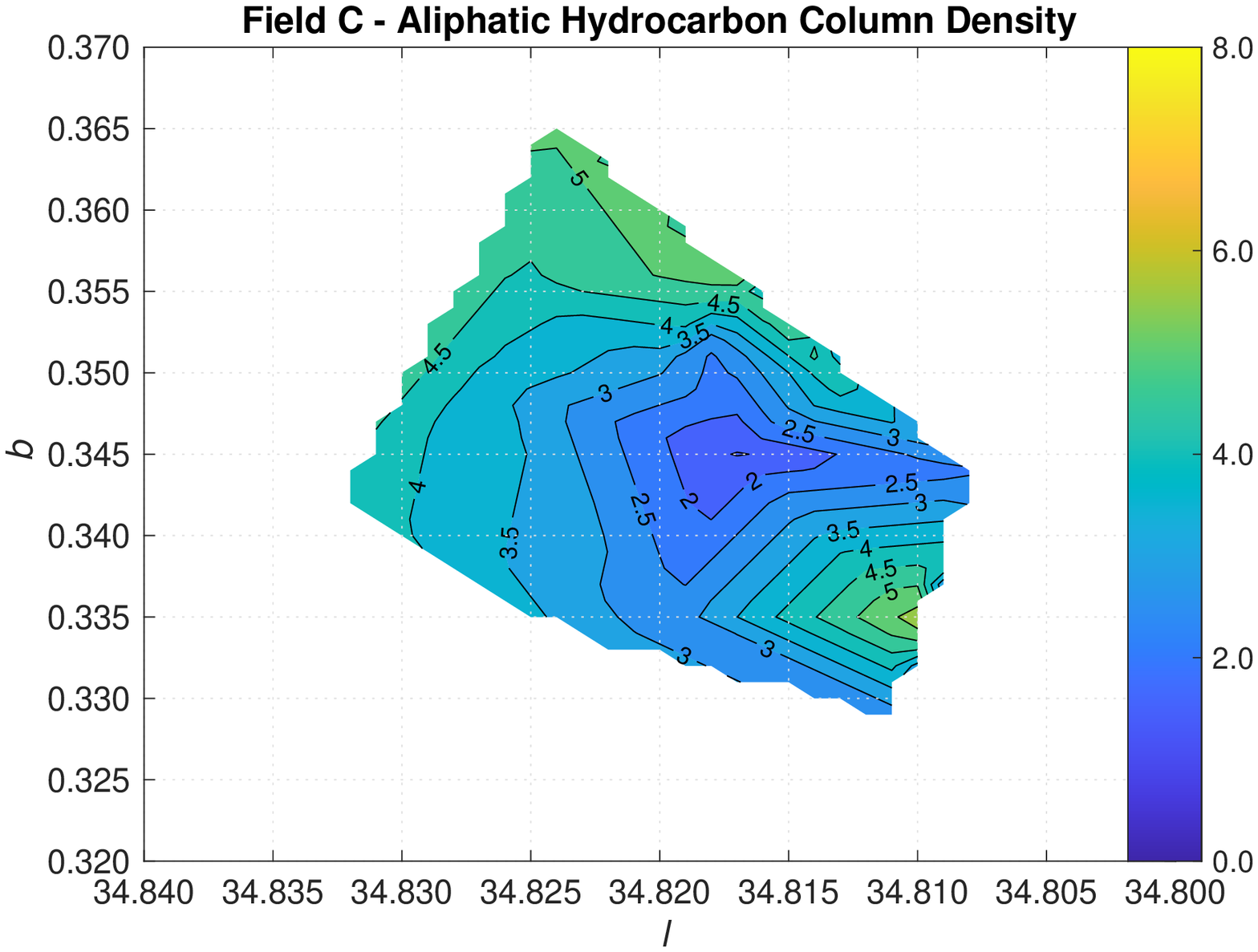}} & {\includegraphics[angle=0,scale=0.46]{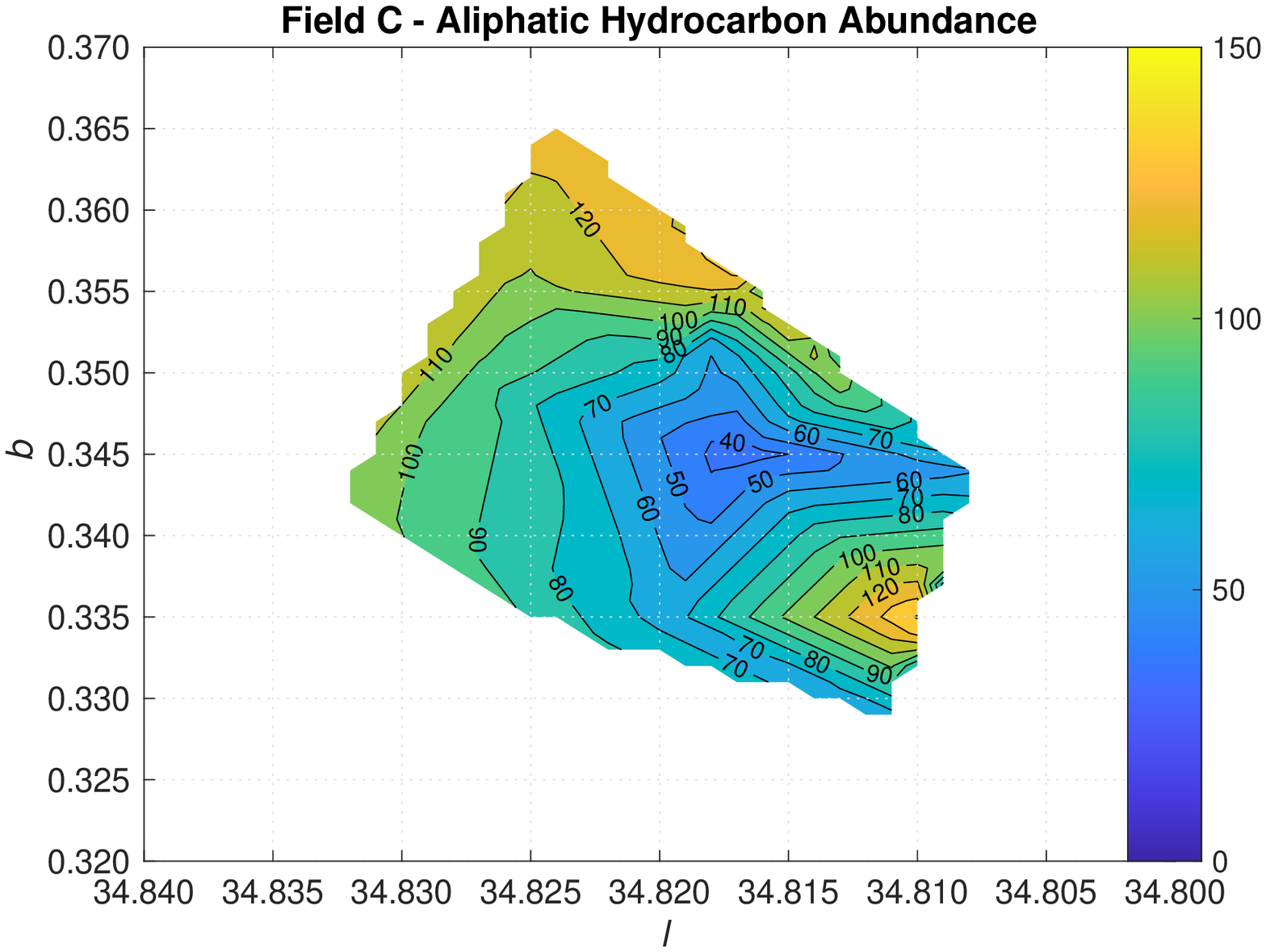}} \\
        \end{tabular}
    \caption{Left panel: maps of aliphatic hydrocarbon column density in galactic coordinates. Colour bars and contours indicate the column density ($\times$10$^{18}$ cm$^{-2}$). Right: maps of the aliphatic hydrocarbon abundance relative to hydrogen (assuming A$_{V}$ is invariant in each field). Colour bar and contours indicate the aliphatic hydrocarbon abundance levels (ppm). Field A, Field B and Field C are presented in upper, middle and below panels respectively. } 
     \label{fig6}
      \end{center}
\end{figure*}

 \section{Results}	\label{Section5}
 
Together with the previous results for Field A, our study of Field B and Field C has brought the largest perspective to date on the amount and distribution of solid phase hydrocarbons. We present the outcomes of this work below.

\subsection{Aliphatic Hydrocarbon Column Densities}

We calculated aliphatic hydrocarbon column densities for Fields B \& C based on the 3.4\,$\mu$m optical depth values in Table~\ref{tab:4} and Table~\ref{tab:5} by applying Eq.~\ref{eq:1} and using the aliphatic hydrocarbon absorption coefficient of \textit{A} = $4.7\times10^{-18}$\,cm\,group$^{-1}$ (as determined in Paper 1). Since optical depths of 3.4\,$\mu$m were obtained by spectrophotometry, we used 3.4\,$\mu$m narrow-band filter width of 62\,cm$^{-1}$, instead of the equivalent width of the 3.4\,$\mu$m absorption feature (108.5\,cm$^{-1}$, see Paper 1) to calculate the aliphatic hydrocarbon column densities. Therefore, the resultant aliphatic hydrocarbon column densities show lower levels than could be detected by spectrophotometry.

The resultant maps of aliphatic hydrocarbon column densities are shown in galactic coordinates for Field B and C in Figure~\ref{fig6} in comparison with the aliphatic hydrocarbon column density map of Field A (Paper 2). We found that there is a rise in hydrocarbon column density through to the Galactic plane in Field B similar to that of Field A.  

\subsection{Relation with ISD Distribution}

For the Galactic Centre fields, to investigate whether there is a similarities between the distribution of the aliphatic hydrocarbons in ISD and the total dust, the maps showing the extinction and/or reddening in visible and near infrared regions have been examined. However, a map with similar resolution and coverage that will show extinction and/or reddening through the line of sight of the GC ($\sim$8 kpc) has not been found in the literature \citep{Sumi2004, Marshall2006, Schodel2007, Schodel2010, Nogueras-Lara2018, Green2019, Chen2019}, as the sightline is obscured due to intervening opaque clouds in visible and near infrared wavelengths and there is lack of photometric data of proper background sources. There are many luminous background sources in the near infrared region, however, the majority of them are cool red giants. 

The GC is totally obscured in the ultraviolet and visible wavelength regions. Thus effects of high extinction and low temperature on IR photometry cannot be easily separated. Therefore highly obscured blue, young massive stars can hardly be distinguished from less obscured red, old low-mass stars \citep{Geballe2010}. 

We tried to probe the dust distribution using maps that are prepared based on thermal emission of ISD in the far infrared region \citep{Schlegel1998, Peek2010, Schlafly2011, PlanckCollaboration2014}, where the extinction can be sufficiently low to enable us to investigate the ISM through the line of sight of the GC. However, we were also unable to find a map with comparable resolution and coverage to analyse interstellar dust distribution at longer wavelengths. 

\begin{figure*}
  \begin{center}
    \begin{tabular}{c}
      {\includegraphics[angle=0,scale=0.46]{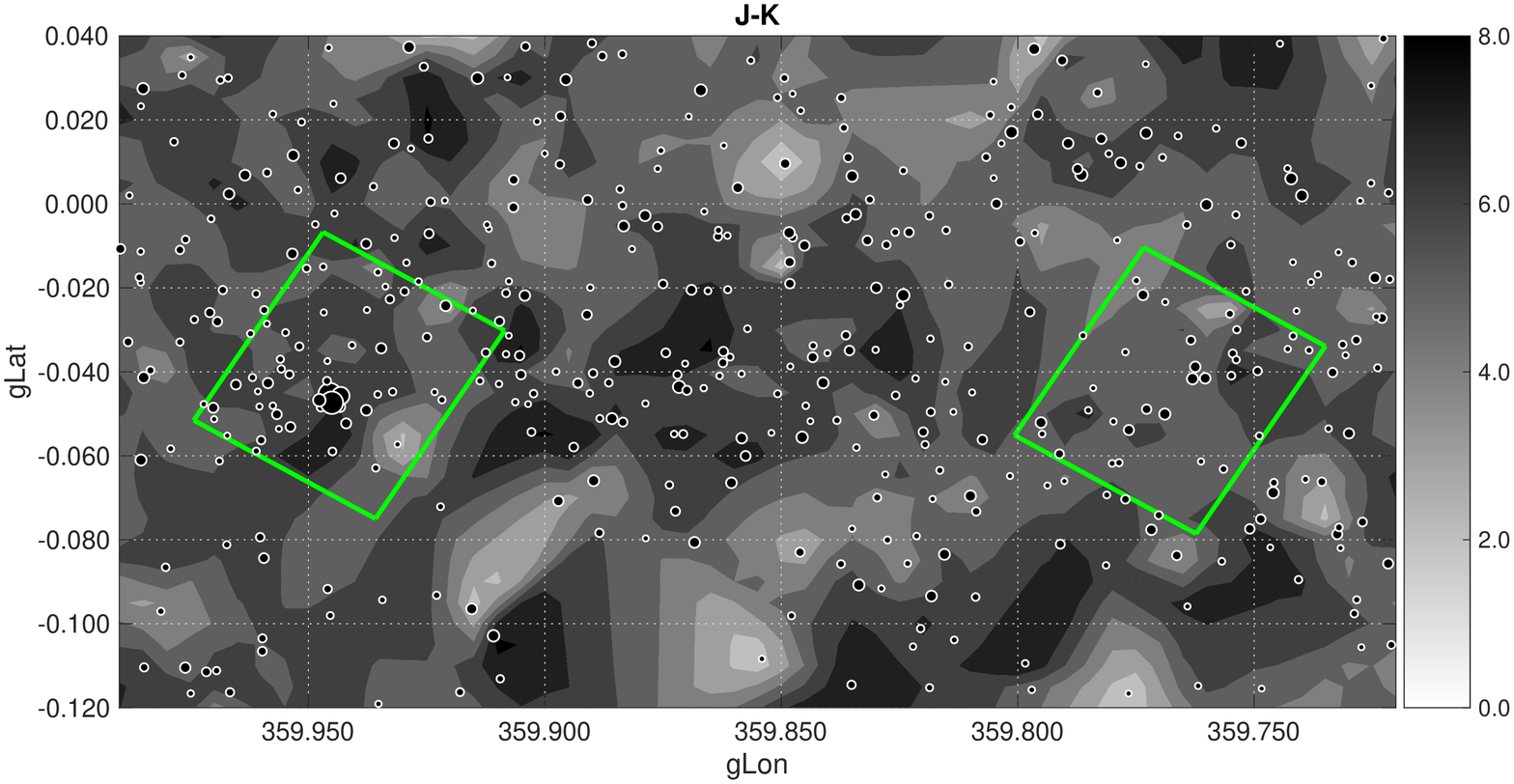}} \\
        {\includegraphics[angle=0,scale=0.46]{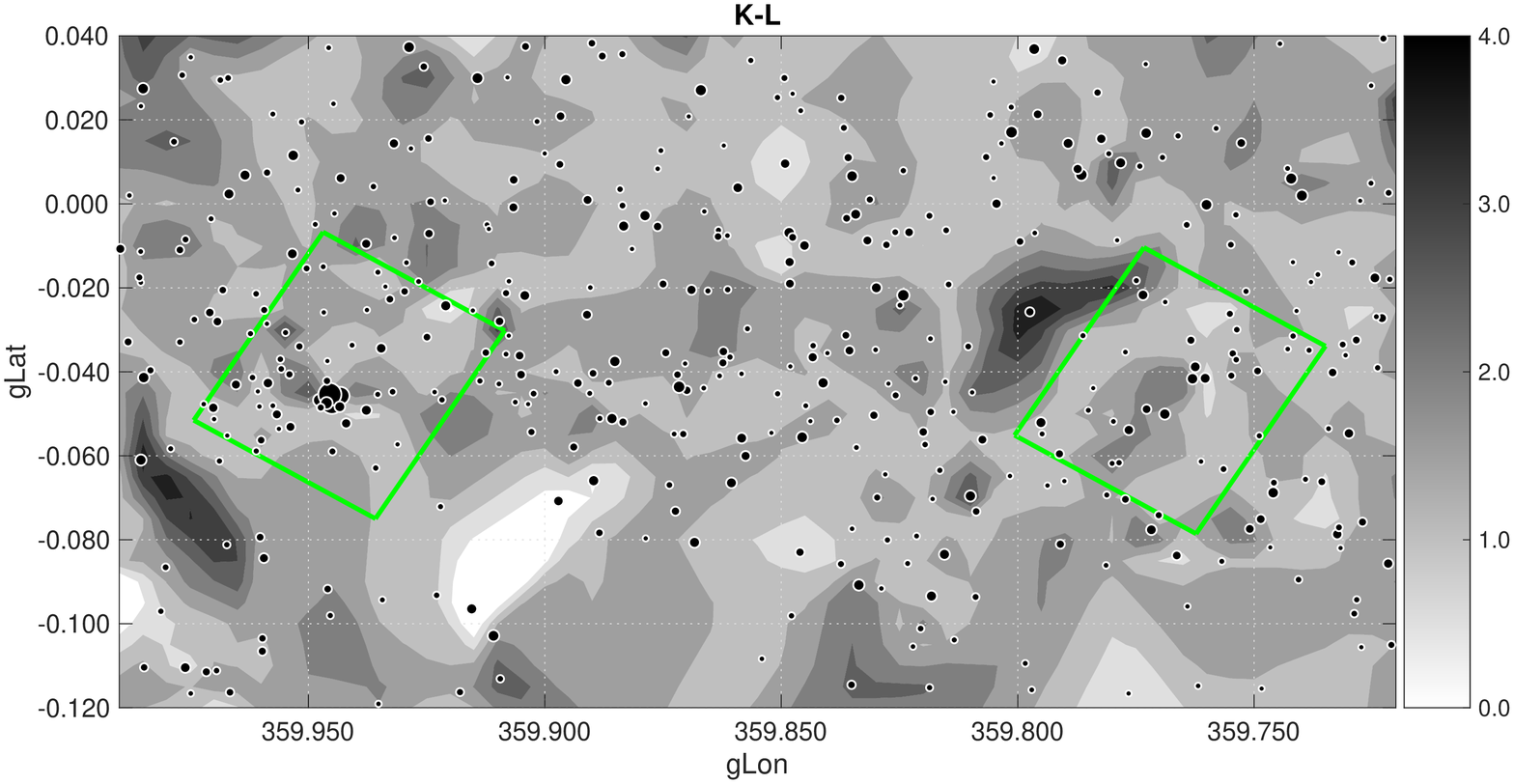}} \\
          {\includegraphics[angle=0,scale=0.46]{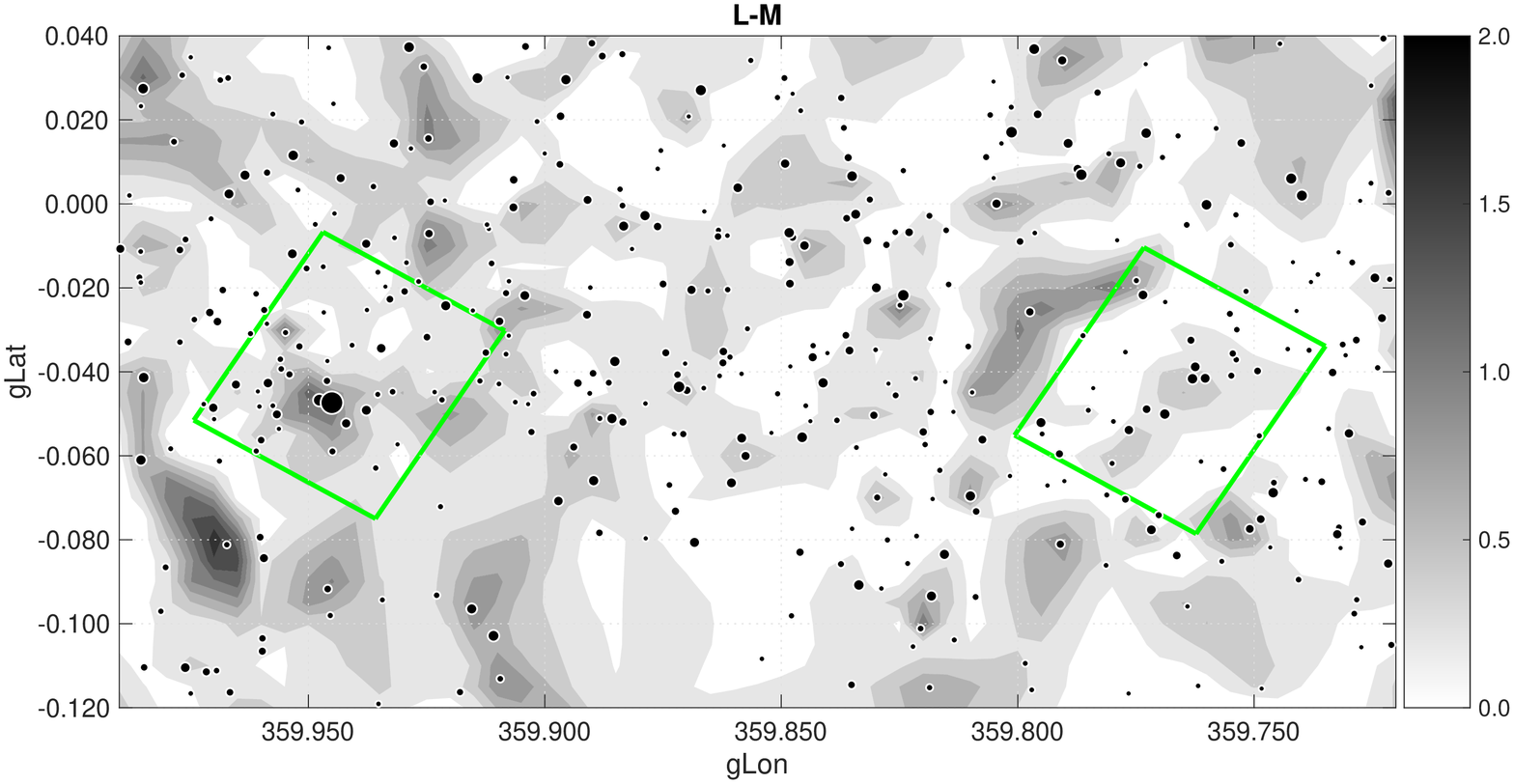}} \\
   \end{tabular}
    \caption{Colour excess maps: 2MASS J$-$K (upper panel), 2MASS/Spitzer K$-$L (Ch1) (middle panel), and Spitzer L$-$M (Ch2) (bottom panel). Field A is on the left and Field B is on the right. Colour bars and contours indicate the colour excess levels in magnitude. }
  
     \label{fig7}
      \end{center}
\end{figure*}

Finally, we decided to analyse the reddening by using the Two Micron All Sky Survey Catalog \citep{Skrutskie2006} and Spitzer GLIMPSE Catalog \citep{Ramirez2007} data sets used in \citealt{Geballe2010}, which were kindly made available by the authors for our use. We prepared colour excess maps that cover the GC fields with a resolution which can enable us to make a meaningful comparison. We used photometric brightness measurements at 2MASS J-band (1.25\,$\mu$m), H-band (1.65\,$\mu$m), K-band (2.17\,$\mu$m) and Spitzer IRAC Ch1-band (3.6\,$\mu$m), Ch2-band (4.5\,$\mu$m), Ch3-band (5.8\,$\mu$m), Ch4-band (8\,$\mu$m) and obtained 2MASS J$-$K, 2MASS/Spitzer K$-$L (Ch1) and Spitzer IRAC Ch1$-$IRAC Ch2 colour excesses of all bright L--band sources (m$_{L}$ \textless 8$^{m}$) in the GC region. We plotted colour excess maps and compared them to check whether there are similar trends with the aliphatic hydrocarbon column density maps. We present 2MASS J$-$K, 2MASS/Spitzer K$-$L (Ch1) and Spitzer L$-$M (Ch2) colour excess maps in Figure \ref{fig7} (note that the colour excess levels of the maps are different: highest for J$-$K and lowest for L$-$M). We also indicate the location of the sources by black dots whose sizes are proportional to their fluxes. However, we cannot determine any relation between the fluxes and colour excesses.

We would like to note that in the colour excess maps presented here, there would be bias due to the extinction and effective temperature degeneracy mentioned above. There are different methods to overcome this degeneracy \citep{Indebetouw2005, Marshall2006, Gonzalez2012, Hanson2014}. However, an extensive study of extinction through the GC sightlines is out of the scope of this study. There is a bias and uncertainty due to circumstellar dust around the mass losing asymptotic giant and ascending giant branch stars. They appear redder and this can be accounted for ISD \citep{Marshall2006, Nogueras-Lara2018}. 

The matching trends between the colour excess maps presented in Figure \ref{fig7}, prove that they sufficiently reflect the distribution of interstellar matter. However, we could not detect any matching trends between the colour excess maps and aliphatic hydrocarbon column density maps.

\subsection{Aliphatic Hydrocarbon Abundances}

We converted our results into aliphatic hydrocarbon abundance ratios, in order to compare with carbon abundances (C/H) reported in the literature. Normalised aliphatic hydrocarbon abundances (ppm) were estimated based on gas-to-extinction ratio $N(H) = 2.04 \times 10^{21}$\,cm$^{-2}$ mag$^{-1}$ \citep{Zhu2017}. 

Average extinction toward the central parsec in the GC is reported to be A$_{Ks}$$\sim$2.5 mag and A$_{V}$$\sim$30 mag \citep{Scoville2003, Nishiyama2008, Fritz2011, Schodel2010}. \cite{RiekeLebofsky1985} found A$_{V}$ to A$_{K}$ ratio is 9, but more recent studies indicate even higher values of   A$_{V}$ to A$_{K}$ ratio is $\sim$ 29 \citep{Gosling2006}. Extinction towards the GC also varies on scales of arcseconds \citep{Scoville2003, Schodel2007, Schodel2010}. However, the extinction maps in the literature have very different resolutions or coverage or both, which therefore limit their use in obtaining the gas density distribution in each field. Therefore, we assumed extinction is invariant to estimate normalised aliphatic hydrocarbon abundances.

For Field A, previously we assumed A$_{V}$$\sim$30 mag (see Paper 1 and Paper 2). In this study, we followed the same approximation for Field B as it is in the GC region. 

Since there is a debate on the extinction towards Field C, to estimate the lowest abundance level, we assumed A$_{V}$$\sim$20 mag, which is the highest value for the DISM reported by \cite{Godard2012} and \cite{Vig2017}. 

The resultant aliphatic hydrocarbon abundance levels (ppm) that are estimated based on an invariant A$_{V}$ are shown in galactic coordinates for Field C and B in Figure~\ref{fig6} in comparison with Field A (for details see Paper 2).

\subsection{Summary}

For completeness, we summarise and compare the results for Fields A, B \& C in Table \ref{tab:7}, which lists average, minimum and maximum 3.4\,$\mu$m optical depth values, aliphatic hydrocarbon column densities and abundances for each field, together with an estimate of the fraction of the aliphatic carbon in the ISM that lies in the ISD\@.

 \begin{table*}
 \begin{center}
 \footnotesize
 \caption{Minimum, maximum and average 3.4\,$\mu$m optical depth ($\tau_{3.4\,\mu m}$) values, with the corresponding aliphatic hydrocarbon column densities ($\times$\,10$^{18}$ cm$^{-2}$) and abundances (ppm) for Fields A, B and C\@.  For the optical depth values the standard deviations are given. We also note the number of sources measured in each field in the footnote to this Table.}
 \label{tab:7}
 \centering 
 \begin{tabular}{| c | c | c | c | c |  }
 \hline 
&&	Field A	&	Field B	&	Field C	\\
\hline 
&Min. &	0.07	&	0.18	&	0.10	\\
Optical depth	($\tau_{3.4\,\mu m}$) & Max. &	0.43	&	0.62	&	0.44	\\
& Average &0.20	&	0.36	&	0.27	\\
& Sigma & 0.06 & 0.09 & 0.10 \\
\hline 
& Min.	&	0.92	&	2.40	&	1.36	\\
Column density	($\times$\,10$^{18}$ cm$^{-2}$ ) & Max. &	5.67	&	7.16	&	5.83	\\
& Average &	2.64	&	4.78	&	3.56	\\
\hline 
& Min. &	15	&	39	&	33	\\
Abundance	(ppm) & Max. &	93	& 117 & 143	\\
& Average 	&	43	&	78	&	87	\\
\hline 	
Aliphatic C ($\%$)&&	12	&	22	&	24	\\

\hline

\end{tabular}
\end{center}
\begin{footnotesize}
Note that number of sources measured in Field A, B and C are as follows: 200, 180, 15.  
\end{footnotesize}
\end{table*}

For Field A, the optical depth at 3.4\,$\mu$m ranges from 0.07 to 0.43, the column density of aliphatic hydrocarbon rises from $\sim$$9.2\times10^{17}$\,cm$^{-2}$ to $\sim$$5.7\times10^{18}$\,cm$^{-2}$, and the corresponding abundances with respect to hydrogen range from $\sim$15\,ppm to $\sim$93\,ppm. The mean value of $\tau_{3.4\,\mu m}$ $\sim 0.2$ corresponds to a typical column density of aliphatic hydrocarbon of $2.6\times10^{18}$\,cm$^{-2}$ and 43\,ppm aliphatic hydrocarbon abundance. 

For Field B, the optical depth at 3.4\,$\mu$m ranges from 0.18 to 0.54, the column density of aliphatic hydrocarbon rises from $\sim$$2.4\times10^{18}$\,cm$^{-2}$ to $\sim$$7.2\times10^{18}$\,cm$^{-2}$, and the corresponding abundances with respect to hydrogen range from $\sim$39\,ppm to $\sim$117\,ppm. The mean value of $\tau_{3.4\,\mu m}$ $\sim$ 0.36 corresponds to a typical column density of aliphatic hydrocarbon of $4.8\times10^{18}$\,cm$^{-2}$ and 78\,ppm aliphatic hydrocarbon abundance. 

For Field C, the optical depth at 3.4\,$\mu$m ranges from 0.10 to 0.44, the column density of aliphatic hydrocarbon rises from $\sim$$1.4\times10^{18}$\,cm$^{-2}$ to $\sim$$5.8\times10^{18}$\,cm$^{-2}$, and the corresponding abundance with respect to hydrogen ranges from $\sim$33\,ppm to $\sim$143\,ppm. The mean value of $\tau_{3.4\,\mu m}$ $\sim$ 0.27 corresponds to a typical column density of aliphatic hydrocarbon of $3.6\times10^{18}$\,cm$^{-2}$ and 87\,ppm aliphatic hydrocarbon abundance. 

The final line of Table \ref{tab:7} lists the percentage of the total carbon abundance in aliphatic form under the assumption that the total carbon abundance for the ISM is 358\,ppm \citep{Sofia2001}. 

We obtain an average 43 ppm, 78 ppm and 87 ppm aliphatic hydrocarbon levels for Field A, Field B and Field C, respectively. In addition to gas phase abundances, these amounts correspond to 12\%, 22\% and 24\% of the cosmic carbon abundances. We conclude that at least 10--20\% of the carbon along these sightlines in the Galactic plane lies in aliphatic form.

\section{Discussion and Conclusions} \label{Section6}

By this work we have clarified that for three fields in the Galactic disk, an important part of cosmic carbon is in aliphatic hydrocarbon form in the ISD. We also showed that the spectrophotometric method is applicable to other fields in the Galactic plane.

We showed in Paper 2 that optical depth at 3.4\,$\mu$m for a field in the Galactic Centre (Field A) can be reliably measured using flux measurements made through a series of narrow band filters that are able to sample the spectrum. We have applied this method to two new fields: $\sim 0.2^{\circ}$ to the East (Field B) and $\sim 35^{\circ}$ to the West (Field C) lying in the Galactic plane. In all cases optical depths were able to be measured and found to lie in the range $\sim 0.1 - 0.6$.

We found larger optical depth through the line of sight of Field B compared to Field A, based on the calibration fluxes we obtained by the spectroscopic measurements of a source (GCIRS 7) in Field A. However, caution needs to be taken since the sources in Field B lack spectroscopic information. We also found that, through the line of sight of Field C (IRAS 18511+0146 cluster), there is a larger optical depth than Field A. This result was checked by cross-calibration tests by using the spectroscopic fluxes of Field C sources from the literature to calibrate Field A data which yields consistent results. 

Complete consistency cannot be expected since the spectroscopic studies involve different steps that we were not able to implement in this study, such as polynomial continuum fitting (which requires higher resolution spectra) or normalisation of the spectra with the star's blackbody curve (which requires to measure temperature of the all sources in the field of view). Since we applied linear continua to low-resolution spectra, which are smoother than the real spectrum, our results may yield minimum values for the optical depth levels of 3.4$\,\mu$m absorption. 

The 3.4$\,\mu$m optical depths are measured simultaneously for each field so the relative variability of column densities for each field are more reliable than the measurements reported by different spectroscopic studies in the literature (for more details see Paper 2). 

The other advantage of the method is that it allows us to trace abundances through the large FoV, compared to single point or long-slit spectroscopy, which is not sufficient to provide spatial information for the FoV. Although recently \textit{integral field spectroscopy} (IFS) has become an important alternative to long-slit spectroscopy, the narrow-band imaging can be still preferred to obtain spatially resolved spectral data for larger FoVs. Using the spectrophotometric method, we measured the optical depth across the fields that covers 163 arcsec $\times$ 163 arcsec for Field A and B, 137 arcsec $\times$ 137 arcsec for Field C. While the range in extrema between minimum and maximum values measured for 3.4\,$\mu$m optical depth may appear relatively large (for Field A from 0.07 to 0.43, for Field B from 0.18 to 0.62 and for Field C from 0.10 to 0.44), the majority of sources are all within 0.1 of the mean 3.4\,$\mu$m optical depth measured in their field. 

For both Fields A and B there is a mild gradient in the optical depth of 3.4$\,\mu$m running in same direction across the 2 arcmin field, rising by about a factor of 50\% in increasing towards $b=0^{\circ}$. The pattern is similar between these fields. There are too few sources measured in Field C, however, to draw any conclusions as to whether a gradient exists across this field.

As we showed by dividing data into quartiles for the Galactic Centre fields, where the number of the sources are sufficiently large, the general optical depth trends in each quartile are found to be consistent. Although interpolation of data caused some structure variations in the quartile maps due to location of the sources in use, without large source-to-source variations, 3.4\,$\mu$m optical depths are found to be reasonably uniform across the fields. However, in case of large uncertainty in the spectrophotometric measurements of individual sources due to low S/N and uncertainties in continuum determination \citep{Chiar2002, Moultaka2004, Godard2012} or presence of intrinsic spectral properties through some lines of sight there might be source-to-source variations. There might be also bias in the measurements since some of the GC sources are known that they have thick circumstellar dust shells \citep{Roche1985, Tanner2005, Viehmann2005, Pott2008, Moultaka2009} and there are also many sources that are problematic to classify \citep{Buchholz2009, Hanson2014, Dong2017, Nogueras-Lara2018}. However, the uniformity in our measurements suggests that the aliphatic hydrocarbon absorption is dominated by material distributed along the sightline in DISM rather than local to each source, and so our method measures large-scale properties of the foreground molecular medium.

We wanted to compare the resultant aliphatic hydrocarbon optical depth maps obtained in this study with the maps in the literature. \cite{Moultaka2005} and \cite{Moultaka2015} provided the map of the optical depth of 3.4\,$\mu$m absorption feature for the GC sightline. However, the coverage of the 3.4\,$\mu$m absorption maps are considerably smaller than the maps obtained in this study. Therefore a meaningful comparison is not possible. They also argued that there is a residual 3.4\,$\mu$m absorption produced by the local medium of the central parsec of the GC. Through the line of sight of the GC, different components of the ISM (diffuse and dense ISM, circumnuclear ring, mini spiral etc.) are superimposed \citep{Muzic2007, Moultaka2009, Ferriere2012, Sale2014, Yusef-Zadeh2017, Moultaka2019, Murchikova2019, Geballe2021}. However, 3.4\,$\mu$m hydrocarbon absorption is assumed to dominantly take place in the DISM \citep{Chiar2002}. Of course, hydrocarbons can be found in all ISM components but by dust evolution, their observable properties change (i.e., ISD could be mixed or covered by ices of the volatiles in dense ISM or in molecular clouds) \citep{Jones2012a, Jones2012b, Jones2012c, Chiar2013, Jones2019, Potapov2021}. There are also masking effects by other features arising from the different components of the ISM and the 3.4$\,\mu$m absorption feature can be superimposed with other features, such as the long wavelength wing of the broad 3.1$\,\mu$m H$_{2}$O ice absorption band ([I02], [C02], [M04] and [G12]).

We also tried to explore if there is a correlation between aliphatic hydrocarbon column density and the ISD by comparing our maps with extinction and reddening maps, although a relation between two cannot necessary be expected. Extinction occurs due to combined effect of absorption and scattering of light and can be used to probe the amount of total gas, ice and dust density together. Although extinction curves are shaped by optical properties of ISD \citep{Draine1984, Cardelli1989, Fitzpatrick1999}, they are not sufficient to reveal to chemical composition of the ISDs. Since our measurement is focused on the 3.4\,$\mu$m feature, the resultant maps reflect distribution of a chemical group: hydrocarbons in the ISD. Importantly, we showed that this distribution is independent from the ISD distribution.   

In addition to the spectrophotometric measurements of the 3.4\,$\mu$m feature, spectrophotometric measurements of the 9.8\,$\mu$m feature has a potential to reveal the siliceous dust distribution but it has not been implemented yet. Spectrophotometric measurement of 3.4\,$\mu$m aliphatic hydrocarbon and 9.8\,$\mu$m silicate absorption features have a potential for the future applications e.g. James Webb Space Telescope (JWST) as discussed in \citealt{Gordon2019}. A comparison of the carbonaceous and siliceous dust maps can help to reveal the abundance distribution of major dust forming elements (i.e. C, O, Si) in the ISM \citep{Kim1996, Cardelli1996, Henning2004, WangLi2015, Zuo2021a, Zuo2021b, HensleyDrain2021, Gordon2021, DraineHensley2021}. 

It has been expected that gas-phase abundances are often assumed to be invariable with respect to the local interstellar conditions although recent studies imply that there are variations \citep{DeCiaNature2021, Zuo2021a, Zuo2021b}. There is an argument by \cite{DeCiaNature2021} on the presence of line-of-sight inhomogeneities in elemental abundances, and measured column densities could be affected by the ISM being composed of individual clouds with very different depletion strengths and/or abundances. In this study, the aliphatic hydrocarbon optical depth is found to be a little higher for Field B than Field A. While this variation occurs within $0.2^{\circ}$ in the centre of the Galaxy, the optical depth in Field C in the Galactic plane is similar to that of the Galactic Centre. While three fields is, of course, a limited number to be drawing conclusions from, this is consistent with reasonably constant levels of absorption at 3.4\,$\mu$m across the Galactic plane.

The resultant aliphatic hydrocarbon column densities have been obtained free from the extinction values. For the three fields, the average aliphatic hydrocarbon column density level found to be several $\rm \times 10^{18}\,cm^{-2}$. We obtained the aliphatic hydrocarbon abundances of several $\rm tens \times 10^{-6}$ (ppm) accordingly for these sightlines. The statements also assume a relatively constant extinction and total carbon abundance, and so caution needs to be applied in order-interpreting it.  In addition, there might possibly be larger amounts of aliphatic hydrocarbons in the ISD as we used 3.4\,$\mu$m narrow-band filter width of 62\,cm$^{-1}$ instead of the $3.4\,\mu$m aliphatic hydrocarbon absorption feature equivalent width of 108.5\,cm$^{-1}$ (Paper 1). Therefore we conclude that an overall fractional abundance of carbon in the aliphatic form from 10--20\% in the ISM\@ at least. 

We also note that the cosmic carbon abundances are still in debate as different studies are not fully consistent yet \citep{DraineHensley2021, Zuo2021a, Zuo2021b}. Using spectroscopic measurements through the different Galactic sightlines, \cite{Parvathi2012} found higher carbon abundance levels (i.e. $\sim$$464$\,ppm towards HD206773) than the cosmic carbon abundance estimations. However, \cite{DeCiaNature2021} implied that the interstellar abundances of refractory elements in the local ISM may be $\sim$55\% of solar, which puts new limitations on elemental abundances in ISD. 

Importantly, the average aliphatic hydrocarbon abundances found in this study do not exceed the solid phase carbon abundance levels obtained by recent studies (i.e. \citealt{Parvathi2012, Jones2013, Mishra2017, Zuo2021b}) and imply that some of the solid carbon is available in aromatic, olefinic and other forms in ISD to produce all observable features \citep{Jones2013}, in particular the 2175\,\AA\, bump \citep{Stecher1965}, which is the strongest extinction feature produced by electronic transitions in carbonaceous material \citep{Mathis1977, Kwok2009, Li2019, Dubosq2020, Xing2020}. By the support of laboratory studies which allow us to estimate the aliphatic hydrocarbon / total carbon ratios in the ISD, the $3.4\,\mu$m aliphatic hydrocarbon maps can play an important role in solving the carbon crisis, understanding interstellar carbonaceous material cycle and chemical evolution of the Galaxy.

The interstellar matter cycle provides the raw material for the formation of stars and planets \citep{Oberg2021}. Stars and planets are formed deep inside dense clouds where siliceous and carbonaceous ingredient of dust covered by ices \citep{Jones2016a, Jones2016b, Oberg2021, Potapov2021}. The gravitational collapse of an interstellar cloud led to the formation of a protoplanetary disks \citep{Andrews2020, Oberg2021} and dust and ice plays role in the formation of the planetesimals (planets, asteroids, and comets) as they collide and stick \citep{Weidenschilling1980}. Assuming the carbon-to-silicon abundance ratio of the the solar photosphere and the ISM is similar (C/Si $\sim$10) \citep{Anderson2017, Asplund2021}, we can estimate that there would be an important amount of aliphatic hydrocarbon available during the planetesimal formation stage. Therefore, beside the role of siliceous / mineral dust and ice (i.e. \citealt{SalterFraser2009, Fraser2010, HillFraser2015, Demirci2019, Musiolik2021}), the possible role of organic material and aliphatic hydrocarbons in dust aggregation and pebble formation \citep{DominikTielens1997, Kazuaki2019, Bischoff2020, Anders2021} which leads to planetesimal formation need to be further investigated using primordial material in the planetesimal samples and analogue materials \citep{schmidtgunay2019} in laboratory experiments.
  
Hydrocarbon groups in ISD are useful to trace the reservoir of organic material and so that prebiotic molecules in the ISM. Some part of the prebiotic molecules have been likely preserved in carbonaceous dust in the planetesimal formation regions \citep{Ehrenfreund2010, Kwok2016, vanDishoeck2020, Ehrenfreund2021, Oberg2021}. The chemical compositions of the planet-forming disks determine hospitality to life \citep{Bergin2015, Oberg2021}. With advent of the new techniques and telescopes, we are able to observe the planet-forming disks and galaxies with great resolution. Future applications of our method will enhance our understanding of the distribution of carbonaceous dust, organic and prebiotic material in space.

 \section*{Acknowledgments}

We would like to thank Dr. Tom Geballe and Dr. Takeshi Oka for their support and for supplying data.

BG would like to thank to The Scientific and Technological Research Council of Turkey (T\"{U}B\.{I}TAK) for their support in this work through the 2214/A International Research Fellowship Programme. TWS is supported by the Australian Research Council (CE170100026 and DP190103151). The University of New South Wales (UNSW) seeded this work through the award of a Faculty interdisciplinary grant.

We also wish to thank the staff at the UKIRT telescope for their help in gathering the data used for this paper through their service programme, in particular Watson Varricatt and Tom Kerr who undertook the observations.  

This research has made use of the NASA/IPAC Infrared Science Archive, which is funded by the National Aeronautics and Space Administration and operated by the California Institute of Technology.

\section*{Data Availability}
The data used in this study will be made available by the corresponding authors upon request.

 \bibliographystyle{mn2e}{}
 \bibliography{List}

\end{document}